\documentclass[twocolumn]{aastex7} 
\DeclareRobustCommand{\ion}[2]{\textup{#1\,\textsc{\lowercase{#2}}}}

\hypersetup{linkcolor=red,citecolor=blue,filecolor=cyan,urlcolor=blue}

\usepackage{graphicx,color}
\usepackage{amssymb}
\usepackage{amsmath}
\usepackage{url}
\usepackage{natbib}
\usepackage{txfonts}

\newcommand{\telap}{$t_{\mathrm{elapsed}}$}
\usepackage{url}
\usepackage{color} 
\usepackage{hyperref}
   
\newcommand{\FeIX}{{\ion{Fe}{IX}}}

\begin{document}
	
	\title{Heating of Solar Coronal Loops by Converging Unipolar Magnetic Flux: Evidence from SDO Observations and a Radiative MHD Simulation \\}
	
	\author[0000-0001-7817-2978]{Sanjiv K. Tiwari}
	\affiliation{Lockheed Martin Solar and Astrophysics Laboratory, 3251 Hanover Street, Palo Alto, CA 94306, USA}
	\affiliation{Bay Area Environmental Research Institute, NASA Research Park, Moffett Field, CA 94035, USA}
    \affiliation{SETI Institute, 339 Bernardo Ave, Mountain View, CA 94043, USA}
    \email{tiwari@lmsal.com}

	\author[0000-0002-2180-1013]{Souvik Bose}
\	\affiliation{Lockheed Martin Solar and Astrophysics Laboratory, 3251 Hanover Street, Palo Alto, CA 94306, USA}
	\affiliation{SETI Institute, 339 Bernardo Ave, Mountain View, CA 94043, USA}
    \affil{Rosseland Centre for Solar Physics, University of Oslo, P.O. Box 1029 Blindern, NO–0315 Oslo, Norway}
	\affil{Institute of Theoretical Astrophysics, University of Oslo, P.O. Box 1029 Blindern, NO–0315 Oslo, Norway}
 \email{sbose@seti.org}

    \author[0000-0002-0333-5717]{Juan Martinez-Sykora}
  \affiliation{Lockheed Martin Solar and Astrophysics Laboratory, 3251 Hanover Street, Palo Alto, CA 94306, USA}
  \affiliation{SETI Institute, 339 Bernardo Ave, Mountain View, CA 94043, USA}
    \affil{Rosseland Centre for Solar Physics, University of Oslo, P.O. Box 1029 Blindern, NO–0315 Oslo, Norway}
	\affil{Institute of Theoretical Astrophysics, University of Oslo, P.O. Box 1029 Blindern, NO–0315 Oslo, Norway}
 \email{jmsykora@seti.org}
  
    \author{Esther Cao}
	\affiliation{Lockheed Martin Solar and Astrophysics Laboratory, 3251 Hanover Street, Palo Alto, CA 94306, USA}
 \email{esthercao888@gmail.com}
    
	\author[0000-0001-7620-362X]{Navdeep K. Panesar}
	\affiliation{Lockheed Martin Solar and Astrophysics Laboratory, 3251 Hanover Street, Palo Alto, CA 94306, USA}
	\affiliation{SETI Institute, 339 Bernardo Ave, Mountain View, CA 94043, USA}
 \email{panesar@lmsal.com}

	\author[0000-0002-5691-6152]{Ronald L. Moore}
	\affiliation{NASA Marshall Space Flight Center, Mail Code ST 13, Huntsville, AL 35812, USA}
	\affiliation{Center for Space and Aeronomic Research, The University of Alabama in Huntsville,
		320 Sparkman Drive,	Huntsville, AL 35805, USA}
 \email{ronald.l.moore@nasa.gov}
    
	\author[0000-0002-8370-952X]{Bart De Pontieu}
	\affiliation{Lockheed Martin Solar and Astrophysics Laboratory, 3251 Hanover Street, Palo Alto, CA 94306, USA}
	\affil{Rosseland Centre for Solar Physics, University of Oslo, P.O. Box 1029 Blindern, NO–0315 Oslo, Norway}
	\affil{Institute of Theoretical Astrophysics, University of Oslo, P.O. Box 1029 Blindern, NO–0315 Oslo, Norway}
 \email{bdp@lmsal.com}
    
	\author[0000-0003-0975-6659]{Viggo H. Hansteen}
	\affiliation{Lockheed Martin Solar and Astrophysics Laboratory, 3251 Hanover Street, Palo Alto, CA 94306, USA}
	\affiliation{SETI Institute, 339 Bernardo Ave, Mountain View, CA 94043, USA}
	\affil{Rosseland Centre for Solar Physics, University of Oslo, P.O. Box 1029 Blindern, NO–0315 Oslo, Norway}
	\affil{Institute of Theoretical Astrophysics, University of Oslo, P.O. Box 1029 Blindern, NO–0315 Oslo, Norway}
 \email{v.h.hansteen@astro.uio.no}

\begin{abstract}
Solar coronal loop heating remains one of the greatest mysteries in astrophysics.
In emerging flux regions, coronal loops form and are heated as strong magnetic flux tubes emerge through the photosphere.
The loop footpoints gradually separate.
In this study, we investigate the heating of quiet-region (QR) coronal loops whose opposite-polarity footpoints are already separated.
Our observations reveal that convection-driven magnetic flux convergence at a loop's foot—where smaller unipolar photospheric magnetic flux patches merge into a stronger flux patch—correlates with loop brightening, and flux divergence at that foot correlates with that loop's disappearance. This behavior is similar to recent findings for solar coronal plumes.
To investigate and quantify this effect, we track 10 QR coronal loops through their lifetimes using SDO/AIA 171 and 94 Å images, along with SDO/HMI magnetograms. 
The coronal EUV luminosity shows a strong correlation with the magnetic flux in at least one footpoint, most strongly when the line-of-sight magnetic field strength (B$_{los}$) exceeds a threshold of $\approx$200–500 G.
A Bifrost MHD simulation of the quiet-Sun network and corona shows similar behavior but at smaller time-scales: magnetic flux convergence and divergence—most pronounced at a loop footpoint—correlate directly with increased and reduced emissivity due to heating and cooling, respectively. 
The simulation reveals an increase in Poynting flux, reconnection, and Joule heating with flux convergence at that footpoint. We hypothesize that an increased rate of nanoflares during flux convergence may explain the enhanced emissivity of these loops due to heating, with potentially broader implications for coronal heating.

\end{abstract}
	

\section{Introduction} \label{sec:intro}

Solar coronal loops are bright arcs of coronal plasma observed in extreme ultraviolet (EUV) emission, tracing closed magnetic field lines anchored in the photosphere and extending into the corona.
Coronal loops are present in both active regions (ARs) and quiet regions (QRs) and exhibit a diverse range of properties depending on their location and formation \citep[see e.g.,][]{webb81,golu99,asch05a,cirt06,klim09,asch11}.
A loop’s lifetime can vary from a few minutes \citep{wine13,tiw19} to several hours \citep{warr02,lope2007}, with many bright loops persisting for 20–40 minutes \citep{pete12,tiw21}. Their footpoint separation distances range from tens to several hundred megameters (Mm) \citep{mulu11}, while their widths have been reported to span from over a Megameter \citep{delz03,asch05a} to ten times smaller than that \citep{pete13,broo13,will20}. While coronal loops are often multi-stranded \citep{schm10} and are frequently observed to be multi-thermal \citep{schm11,nist14,kris17}, their overall temperatures span a broad range, from approximately 10$^5$ K to over 10$^7$ K \citep{asch05,klim06,real14,hino19}.
These observational characteristics suggest that a range of formation and heating mechanisms may govern different types of coronal loops.

The coronal heating problem has attracted numerous studies attempting to match solar observations with theoretical models and hypotheses  \citep{real14,klim15,pont17,saku17,cran19,hino19}. Currently, the foremost theories of loop heating include nanoflares \citep{gold60,parker83a,parker88} and wave heating \citep{alfv47,oste61,tomc07,jess09,vanb11}.	The nanoflare theory hypothesizes that magnetic field lines become entangled by random horizontal convective motions at their feet, resulting in small-scale magnetic reconnection within the loop body. During this reconnection process, magnetic energy is converted into thermal energy, heating the plasma. An individual nanoflare may result in a cascade or storm of fine-scale magnetic reconnections \citep[e.g.,][]{klim09}, making a “subflare” \citep[e.g.,][]{sves76,parker88,moor80,tiw14,moor24}, further channeling heat in the larger-scale braided coronal loops. Nanoflare heating is more suitable for closed coronal loop systems, while the wave heating mechanism may be more appropriate for open field \citep{depo07,cirt13}, or diffuse background \citep{cirt06,tiw23,flei25}. 
In general, establishing whether one or both of these mechanisms work together to heat the solar corona \citep[e.g.,][]{sriv25} remains a great challenge.
 
Coronal loops formed in emerging flux regions are generally bright, hot \citep[up to $\sim$10~MK, see e.g.,][]{2024NatAs...8..706L}, and relatively transient. Strong magnetic-flux tubes generated in the convection zone rise buoyantly, pierce the photosphere, and emerge into the corona exhibiting various plasma dynamics \citep[e.g.,][]{upen25}. Plasma initially embedded in the rising magnetic field can be carried upward, while plasma subsequently drains along the magnetic-field lines toward the lower atmosphere under gravity \citep[e.g.,][and references therein]{cheu14}. In photospheric magnetograms, this formation mechanism is observed as two opposite-polarity flux patches emerging close together, then gradually separating and growing larger as the flux arch rises.
	
Thus, the photospheric magnetic field in which a coronal loop is rooted plausibly impacts, directly or indirectly, the amount of heating the loop has \citep{golu80,webb81,asch01,asga19}. Although scaling laws suggest that loop heating scales directly with the photospheric magnetic field, with the heating rate proportional to a positive power of the field strength, B \citep[e.g.,][]{rosn78,mand00,ugar19}, this relationship does not necessarily hold for all coronal loops. The convective freedom at the loop footpoints, in addition to the field strength B, is also a key factor in determining their heating \citep{tiw17}. \cite{tiw17} showed that the loops connecting sunspot umbrae (both of their feet being rooted in dark sunspot umbrae) are among the dimmest loops of active regions, despite these loops connecting in the strongest field regions on the solar surface, thus challenging the range of validity of scaling laws. This happens likely because stronger magnetic fields suppress convection more effectively \citep[see, e.g.,][and references therein]{apar25}.
		
Magnetic flux coalescence (convergence and merging) and divergence (splitting and dispersal) continuously occur on the solar surface and have long been reported and discussed \citep{mart88,prie94,schr97,fish98,pane16qr,gosi18,pane17,pane18,pane21,moor22,moor23}. However, the role of magnetic flux convergence and divergence in solar coronal loop heating has not been investigated much, except for a couple of research publications adhering to coronal bright points \citep{madj23} and solar plumes \citep{wang16,aval18}. \cite{wang16} observed that, for on-disk solar coronal hole plumes, converging field (driven by converging convection flows) qualitatively correlates with the plume heating in EUV 171 and 193 \AA\ images, and diverging flux density (driven by diverging flows) accompanies the fading away of plumes. \cite{aval18} found a similar result and extended these results to plumes observed in on-disk non-coronal-hole quiet solar regions.  \cite{aval18} further quantified the flux convergence vs plume heating correlation and found that the plume heating strongly correlates with the plume-foot magnetic field strength when the pixels above a certain threshold field strength are used, finding this critical strength to range 200--600 G for different plumes. Due to larger noise in the lateral component of the magnetic field than in the line-of-sight field (B$_{los}$) in SDO/HMI magnetograms \citep{sche12}, \cite{aval18} only used line-of-sight magnetograms. 

Because several locations on the solar surface showed a similar flux convergence and divergence (quantitatively comparable as well), yet no plumes formed there, \cite{aval18} concluded that magnetic flux convergence is a necessary, but not sufficient, condition for plume brightening. To the best of our knowledge of SDO/AIA observations, none of the (on-disk) coronal plumes on the Sun form without flux convergence, and plume decay is always accompanied by magnetic flux divergence. 
However, flux convergence and its correlation with coronal loop heating have not been explored previously. This connection between photospheric magnetic flux convergence and coronal loop heating is the focus of this paper. 

\section{Data and Methods}\label{sec:data}
	
\subsection{Data}
	
	The observational data used in this study were obtained from the Atmospheric Imaging Assembly (AIA) \citep{leme12}, and Helioseismic Magnetic Imager (HMI) \citep{sche12}, both aboard NASA’s Solar Dynamics Observatory (SDO) satellite  \citep{pesn12}. We also analyze, in detail, a synthetic coronal loop computed from the enhanced network simulation by \cite{carl16} using the \textit{Bifrost} code \citep{gudi11} detailed in section \ref{mhdsim}. 
	
	We use SDO/AIA data to track coronal loops over time, as they rotate across the solar disk. AIA has a cadence of 12 seconds at a spatial resolution of 1.5\arcsec\ for full-disk images of the solar corona in seven Extreme Ultraviolet (EUV) passbands, sensitive to temperatures between 0.45 MK and 10 MK \citep{leme12}. We study coronal loops that are well isolated and mainly use the SDO/AIA 171 \AA\ channel, in which coronal loops have the greatest contrast. The 171 \AA\ channel detects primarily \ion{Fe}{ix}\ emission from plasma at temperatures roughly between 0.6~MK and 1~MK. We also use AIA 94 \AA\ passband images to detect hotter plasma in the loops. The AIA 94~\AA\ passband not only captures the emission of \ion{Fe}{XVIII} from plasma at about 6-8 MK, but also captures the emission of plasma at 1~MK \citep[e.g.,][]{warr12}.  
	
	We use SDO/HMI line-of-sight (LOS) magnetograms to track the evolution of the LOS magnetic field (B$_{los}$) at the loop footpoints. HMI provides full-disk LOS magnetograms showing the distribution of photospheric magnetic field at 45~s cadence and 1\arcsec\ spatial resolution.
	To minimize projection effects in magnetic flux measurements \citep{falc16}, as well as from the altitude difference between photospheric flux and coronal features, we study loops near the solar disk center.
    
	The SDO/AIA 171\AA\ data, and HMI magnetograms were downloaded from the Joint Science Operations Center (JSOC) at a 6-minute cadence over the loop’s lifetime, beginning when the flux convergence started, and the loop became visible, and ending upon the loop's disappearance. While the aforementioned higher cadence data are available for AIA and HMI, we use a six-minute cadence for both, which is suitable for our purposes. 
	
    Using the SunPy software package \citep{sunp20}, the AIA and HMI FITS files were converted into standardized maps, and AIA intensity normalization was done by dividing each image by its exposure time. From the normalized AIA images, we measured loop luminosity. Similarly, magnetic field strengths were obtained from HMI B$_{los}$ magnetograms.

\begin{deluxetable*}{cccccccccc}
	\tablenum{1}
	\tablecaption{Specifics of 10 Selected Quiet Sun Coronal Loops}
	\tablewidth{0pt}
	\renewcommand{\arraystretch}{1.0}
	\tablehead{\colhead{Loop} & \colhead{Date and Peak Time} & \colhead{Start Time} & \colhead{End Time\tablenotemark{a}} & \colhead{Sequence Duration} & \colhead{Loop Lifetime} & \colhead{Position\tablenotemark{b}}  & \colhead{Magnetic Flux\tablenotemark{c}} & \colhead{Bz\_thresh\tablenotemark{d}} & \colhead{Loop heating\tablenotemark{e}} \\
    \colhead{Index} & \colhead{(date, UT) } & \colhead{(date, UT) } & \colhead{(date, UT)} & \colhead{ $\sim$hours} & \colhead{$\sim$minutes} & \colhead{(X arcsec, Y arcsec)} & \colhead{Mx} & \colhead{Gauss} & \colhead{nature}  
	}
	\startdata
	1  & 2015 Jul 27, 01:41:33 & Jul 26, 13:40 & Jul 27, 13:40 &  24 & 60 &  $(50, 75)$ & $-$5.1$\times$ 10$^{19}$ & $-$490  & Asymmetric \\
	2  & 2017 Jan 13, 15:53:33 & Jan 13, 10:55 & Jan 13, 20:55 & 10 & 48 &  $(-50, 280)$ & $-1.7$ $\times$ 10$^{19}$ & $-240$  & Asymmetric   \\	
    3  & 2018 Jan 27, 13:05:33 & Jan 27, 09:23\tablenotemark{f} & Jan 27, 15:23 & 6 & 30 & $(280, 345)$ & $-$1 $\times$ 10$^{19}$ & $-$240  & Symmetric\tablenotemark{g}  \\	
    4  & 2018 Jan 28, 01:05:33\tablenotemark{h} & Jan 27, 22:05 & Jan 28, 08:53 & 11 & 66 & $(-200, 320)$ & $2$ $\times$ 10$^{19}$ & $490$  & Symmetric \\
	5  & 2018 Jul 22, 16:14:33 & Jul 22, 11:15 &Jul 22, 20:08 & 9  & 60 & $(-200, 120)$ & $-3.6$ $\times$ 10$^{19}$ & $-440$  & Asymmetric \\
	6  & 2019 Jul 01, 23:11:33 & Jul 01, 09:10 & Jul 02, 09:10 &  24  & 120  & $(-280, 150)$ & $4.6$ $\times$ 10$^{19}$ & $355$  & Symmetric \\
	7  & 2019 Jul 02, 20:14:33 & Jul 02 14:15 & Jul 03, 05:20\tablenotemark{i} & 15 & 72 & $(200,120)$ & $-1.3$ $\times$ 10$^{19}$ & $-215$  & Symmetric  \\
	8  & 2020 Jul 06, 05:59:33 & Jul 06 04:00 & Jul 06, 13:00 & 9 &  60 & $(-225, 330)$ & $2$ $\times$ 10$^{19}$ & $260$  & Symmetric \\
	9  & 2020 Aug 24, 23:29:33 & Aug 24, 11:30 & Aug 25, 11:30 & 24  &  90 & $(-200, -500)$ & $-1$ $\times$ 10$^{19}$ & $-440$  & Asymmetric \\	
	10  & 2020 Sep 02, 18:52:57 & Sep 02, 06:55 & Sep 03, 03:55 & 21  & 54 & $(-400, 370)$ & $3.5$ $\times$ 10$^{19}$ & $250$  & Asymmetric\\
	\enddata	
	\tablenotetext{a}{The start and end times listed here span the period from the onset of magnetic-flux convergence to the end of flux divergence. The duration of this interval varies among the different loop sets, ranging from approximately six to 24 hours, as listed under the column ``Sequence Duration". Please note that the provided animations may sometimes be longer than the duration listed in the Sequence Duration, as they are intended to provide additional context.
    We also note that this interval is substantially longer than the lifetime of any individual coronal loop observed during the corresponding data sequence, which ranges from approximately six to 60 minutes. The lifetime of principal loop ranges from 30 minutes to 120 minutes, as listed in the column ``Loop Lifetime".}
	\tablenotetext{b}{The approximate central position of the box outlining the loop at their peak time.}
    \tablenotetext{c}{Absolute magnetic flux measured at the converging footpoint of the loop during peak flux.}
    \tablenotetext{d}{B$_{los}$ critical value at the converging loop foot above which the correlation between total B$_{los}$   (or B$_z$ for disk center) and loop luminosity is maximum. }
    \tablenotetext{e}{Whether a loop exhibits symmetric heating (starting at the apex or brightening nearly simultaneously along its length), or asymmetric heating that begins at a loop footpoint on the flux-converging side.}
	\tablenotetext{f}{This is a time very close to the loop formation with visible strong enhancement - as captured in the corresponding movie, flux convergence and some intensity enhancement in the loop system start earlier than the start time listed here.}
    \tablenotetext{g}{In this case, flux of both foot/polarities converge, the negative one showing slightly better correlation, and the loop heating is more abrupt and symmetric.}
   \tablenotetext{h}{In this case, two successive principal loops are observed (see the corresponding movie), with the first forming from positive-flux convergence and the second from negative-flux convergence. For consistency, here we list only the timing and characteristics of the first loop.}
   \tablenotetext{i}{Sometimes, magnetic flux converges again after the initial convergence and divergence, causing loops to be repeatedly formed. This is an example of such a case: at 05:20:33 on 2019-07-03, the main loop almost disappears, but then reforms as a result of renewed flux convergence.}
	\label{tab:1}
\end{deluxetable*}

\subsubsection{Loop identification and selection}\label{selection}

We used Helioviewer \citep{mull17_JHelioviewer} to identify isolated coronal loops in quiet-Sun regions using AIA 171 \AA\ and 94 \AA\ observations. For consistency, we have excluded loops located in coronal holes and active regions and restricted our analysis to loops in quiet-Sun regions. For each candidate loop, we examined the temporal evolution of the photospheric magnetic field at its footpoints using SDO/HMI LOS magnetograms. To exclude loops associated with flux-emergence events, we inspected the HMI magnetograms for signatures of emerging flux, particularly the appearance and subsequent separation of opposite-polarity magnetic flux concentrations. Loops for which no clear evidence of such flux emergence was found during the heating phase were retained for further analysis. 

For many of the selected loops, we found a clear temporal correspondence between the evolution of the photospheric magnetic flux and the coronal emission. In particular, convergence of magnetic flux at least at one footpoint was generally accompanied by an increase in loop brightness, whereas flux divergence was associated with loop dimming. Based on the clarity of the coronal structures, the quality of the associated HMI observations, and the apparent correspondence between magnetic-flux evolution and loop emission, we selected ten coronal loops for detailed analysis. Their properties are summarized in Table~\ref{tab:1}.

For the luminosity measurements, we visually inspected the AIA images and defined a rectangular region of interest (ROI) around each loop, primarily using the AIA 171~\AA\ images. The ROI was chosen to be as small as possible while encompassing the entire visible loop, including both footpoints. We then tracked the loop luminosity throughout its lifetime by summing the AIA 171~\AA\ intensity within this ROI at each time step.

\begin{figure*}
		\centering
		\includegraphics[trim=0cm 0cm 0cm 0cm,clip,width=\linewidth]{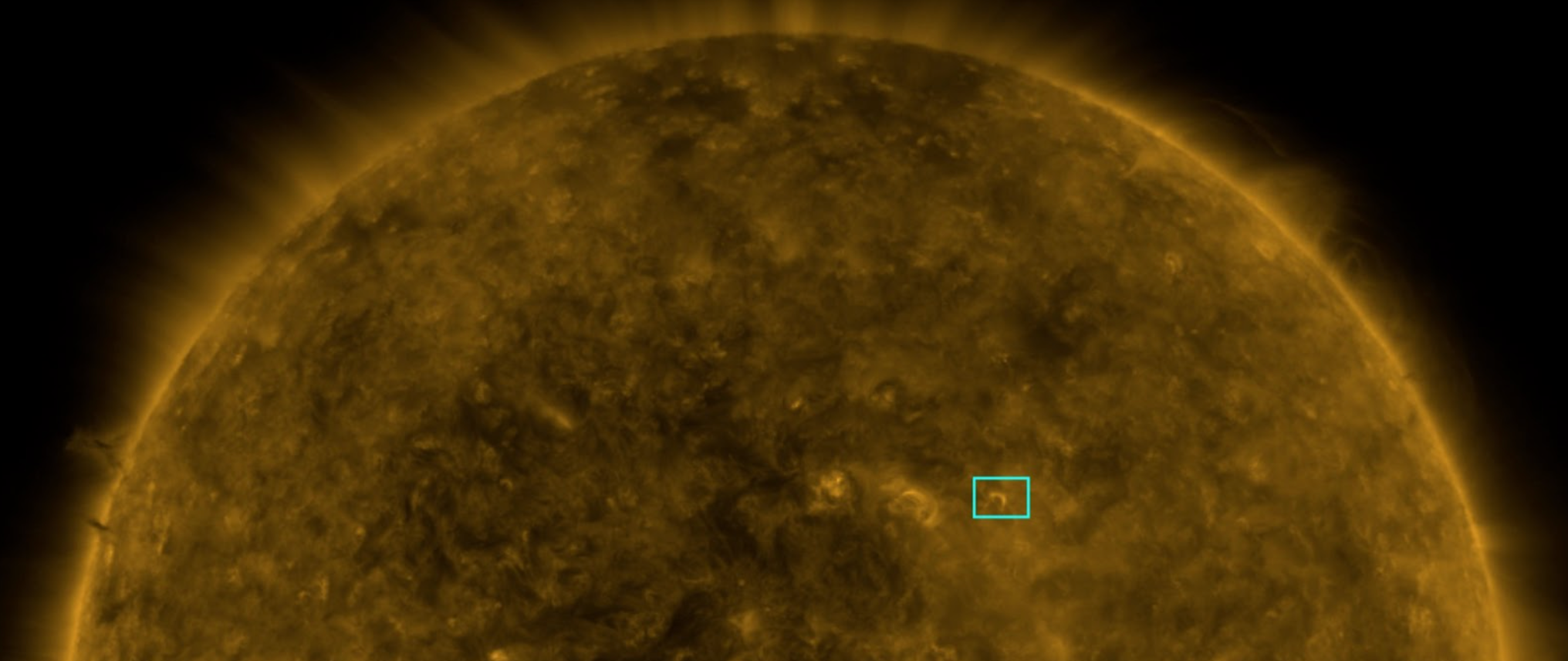}
        \includegraphics[trim=0cm 0cm 0cm 0cm,clip,width=\linewidth]{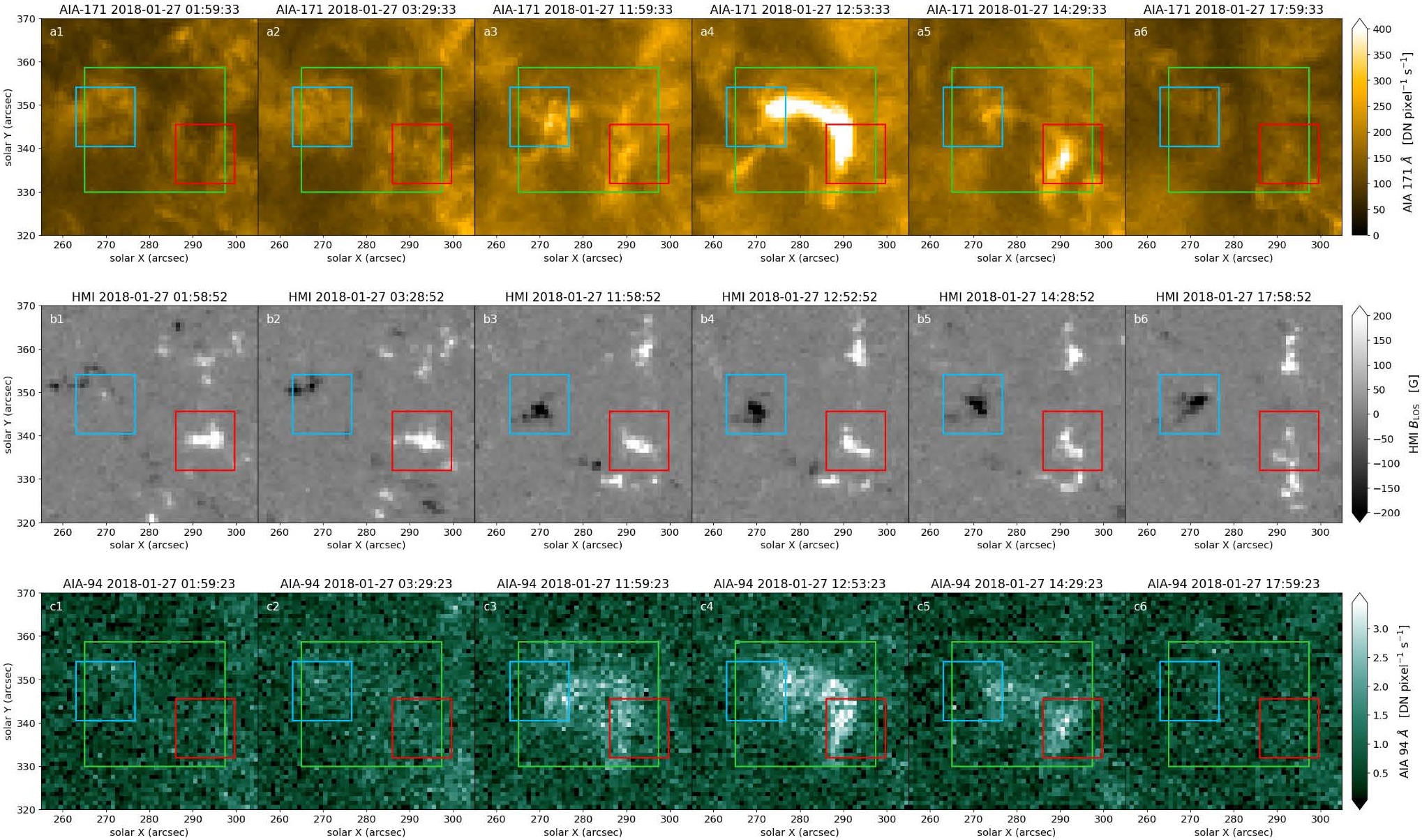}
		\caption{Evolution of loop 3 from Table 1. The top panel shows a large field of view of the Sun in AIA 171 \AA, with a green box outlining the loop in a quiet Sun region. The bottom three rows show the evolution of the loop in AIA 171 \AA, HMI LOS magnetograms, and AIA 94 \AA\ images from top to bottom, respectively. These three panels have color bars on their right sides, consistent with the corresponding images in this figure and with other AIA and HMI images. The cyan and red boxes outline the loop feet, covering negative and positive polarity magnetic flux, respectively. In AIA 171 and 94 \AA\ images, a large green box outlines the area in which the loop luminosity, plotted later in Figure \ref{fig:2_plots}, is calculated. An animation spanning $\sim$25 hours—much longer than the sequence duration listed in Table \ref{tab:1}, for context—with a temporal cadence of six minutes is available online (Movie1\_loop\_index3.mp4). The animation runs for 12 seconds. 
        } 
		\label{fig:1_context}
	\end{figure*}

For the HMI magnetic-field measurements, we similarly identified the relevant ROI by examining the temporal evolution of the HMI LOS magnetograms. By superimposing the HMI and AIA observations, we connected the magnetic-flux concentrations to the observed loop footpoints. We then defined a rectangular ROI around each footpoint that was as small as possible while encompassing the individual magnetic-flux elements that subsequently converged toward the loop footpoint and excluding nearby flux concentrations that did not appear to contribute to the evolution of the loop. The same ROI was maintained throughout the analyzed time interval to track the temporal evolution of the magnetic flux associated with the loop footpoint. Please refer to Figure~\ref{fig:1_context} and Section~\ref{sec:results} for further details and an example.

By magnetic-flux convergence, we mean the coalescence of same-polarity magnetic-flux elements, whereby smaller flux concentrations converge and merge to form larger magnetic elements, as occurs in the photospheric magnetic network. This should be distinguished from the convergence and cancellation of opposite-polarity flux concentrations. In the loops investigated here, we do not observe the latter behavior at the loop footpoints; instead, the two opposite-polarity footpoint regions remain separated. This is distinct from the magnetic evolution commonly associated with coronal bright points, where opposite-polarity flux concentrations converge and cancel  \citep{gals17,madj19,nobr22,tiw22}. 

As discussed in more detail in Section \ref{results_flux_convergence_loop_heating}, the observed loop systems are dynamic and do not necessarily consist of a single, persistent loop throughout their evolution. During periods of photospheric magnetic-flux convergence, multiple small-scale, spatially isolated loop structures can appear and disappear successively, collectively forming the loop systems that we refer to as ``sequence" of coronal loops, see, e.g., in Table~\ref{tab:1}. These loop sequences develop and brighten during magnetic-flux convergence and subsequently fade as the underlying magnetic concentrations diverge and/or decay. Thus, the evolution of the observed coronal structures is closely associated with the evolution of the photospheric magnetic field.

\subsection{Bifrost MHD simulation of an enhanced network region}\label{mhdsim}

We use a publicly available\footnote{\url{https://sdc.uio.no/search/simulations}} radiative MHD simulation of an enhanced network region computed using the Bifrost code to examine whether simulated loops are consistent with the observed flux-convergence-formed loop results \citep[en024048\_hion,][]{carl16}. This examination would help in better interpreting the observational findings. Conversely, the validity of the simulated loops is verified by comparing their properties with the observed loops. Bifrost is a modular, massively parallel, 3D radiation-MHD code \citep[see][]{gudi11} that solves the MHD equations on a staggered grid. The code also includes optically thick radiative transfer in the photosphere and the chromosphere \citep{haye10}, non-LTE radiative losses in the chromosphere \citep[using the recipes from][]{carl12}, and optically thin losses in the transition region and corona. The simulation also includes non-equilibrium ionization effects in hydrogen, following \cite{leen07}, and thermal conduction along the magnetic field lines. For this paper, we have analyzed snapshots 323 to 480 at a cadence of 10~s. 

The numerical domain of the simulation spans 24~Mm~$\times$~24~Mm in the horizontal direction and from $-$2.4~Mm below the $\tau_{500}=1$ surface to 14.4~Mm above it. The horizontal grid size is 48~km, whereas the vertical resolution varies from $19$~km in the photosphere and the chromosphere to about $100$~km in the corona. The net magnetic flux is roughly zero, with two main magnetic fields of opposite polarities separated by $8$~Mm, forming loops $\sim12$~Mm long, mimicking an enhanced network, and an unsigned photospheric area-averaged flux of $46$~G.

To compare with the observations, we synthesized the optically thin \ion{Fe}{IX} emission without absorption in statistical equilibrium from the simulation, which is the dominant contributor to the SDO/AIA 171~\AA\ passband. Since observed loops are most clearly visible in 171 \AA\ images, we use only \ion{Fe}{ix} synthetic data for further investigation. The intensity synthesis was performed by integrating the contribution function $G(T,n_{e})$, obtained from the CHIANTI database \verb |v10| \citep[][]{dere97,2021ApJ...909...38D}, along the LOS using electron density ($n_{e}$), assuming photospheric abundances \citep{aspl09}, local hydrogen density ($n_{H}$), and temperature ($T$). Additionally, we computed the \ion{Fe}{ix} emissivity ($n_{e}\times n_{H}\times G(T,n_{e})$) to trace the emission along the magnetic field lines forming the synthesized loops. Such an analysis provided a convenient way to correlate flux convergence (divergence) with the enhancement (dimming) of emissivity along the simulated loops. 

	\begin{figure*}
		\centering
		\includegraphics[trim=2cm 32cm 15cm 1cm,clip,width=\linewidth]{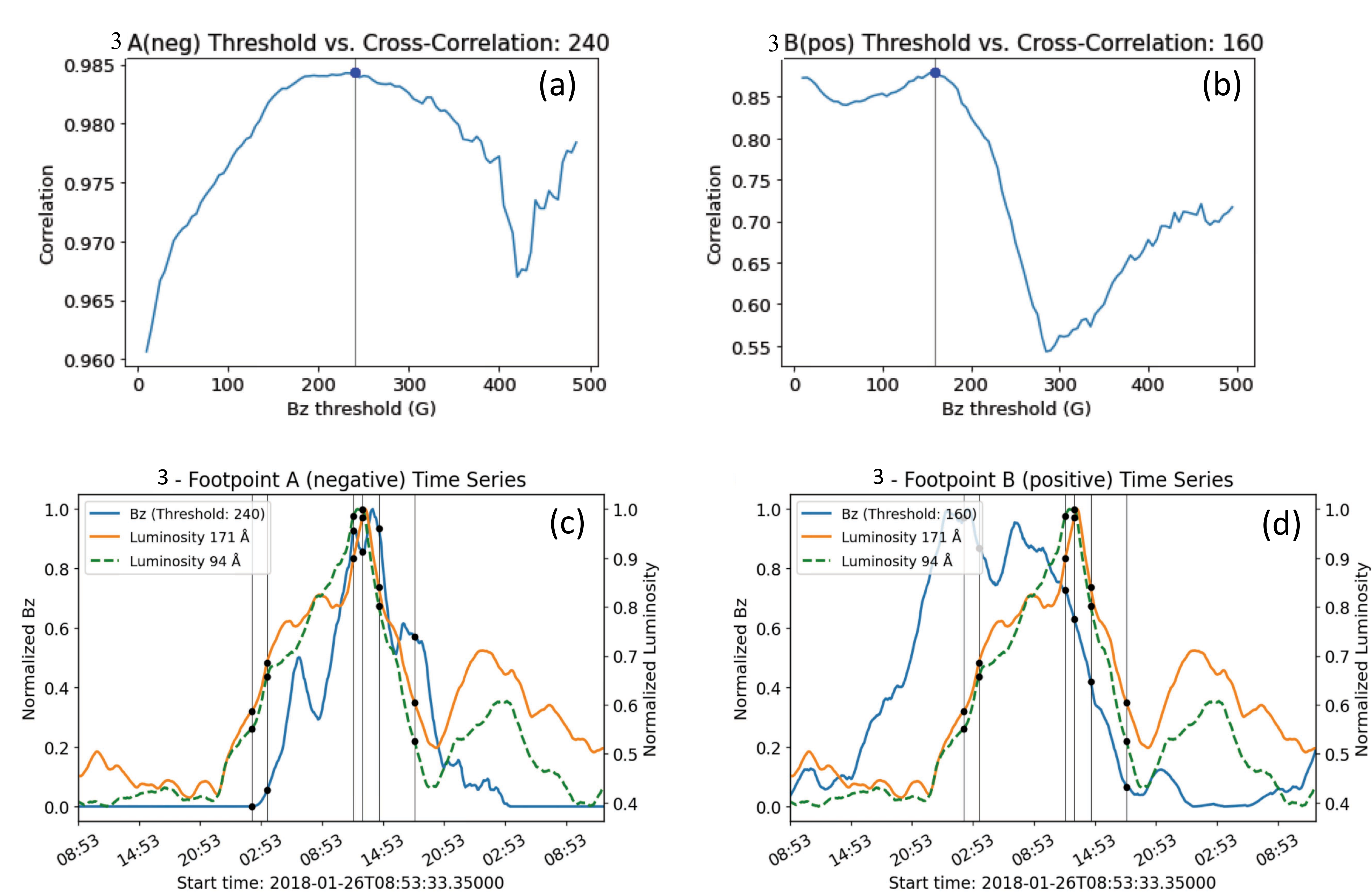}
		\caption{Different plots corresponding to the loop 3 shown in Figure \ref{fig:1_context}. The negative polarity foot is named A, and the positive polarity foot is named B in the title of the plots.  (a) Cross-correlation coefficients estimated between the loop luminosity time profiles and the LOS magnetic field time profiles for the loop foot A plotted for threshold B$_{los}$ values, ranging from 0 to 500 G. (b) same as (a) but for the loop foot B. (c)  The luminosity plots in AIA 171 and 94 \AA\ for loop foot A, with the total B$_{los}$ over-plotted for the threshold of 240 G estimated from panel (a). The vertical six lines mark the timing of six images shown in the last three rows of Figure \ref{fig:1_context}. (d) same as (c) but for the loop foot B. 
		} 
		\label{fig:2_plots}
	\end{figure*}

	\begin{figure}
		\centering
		\includegraphics[trim=33cm 0cm 64cm 0cm,clip,width=\linewidth]{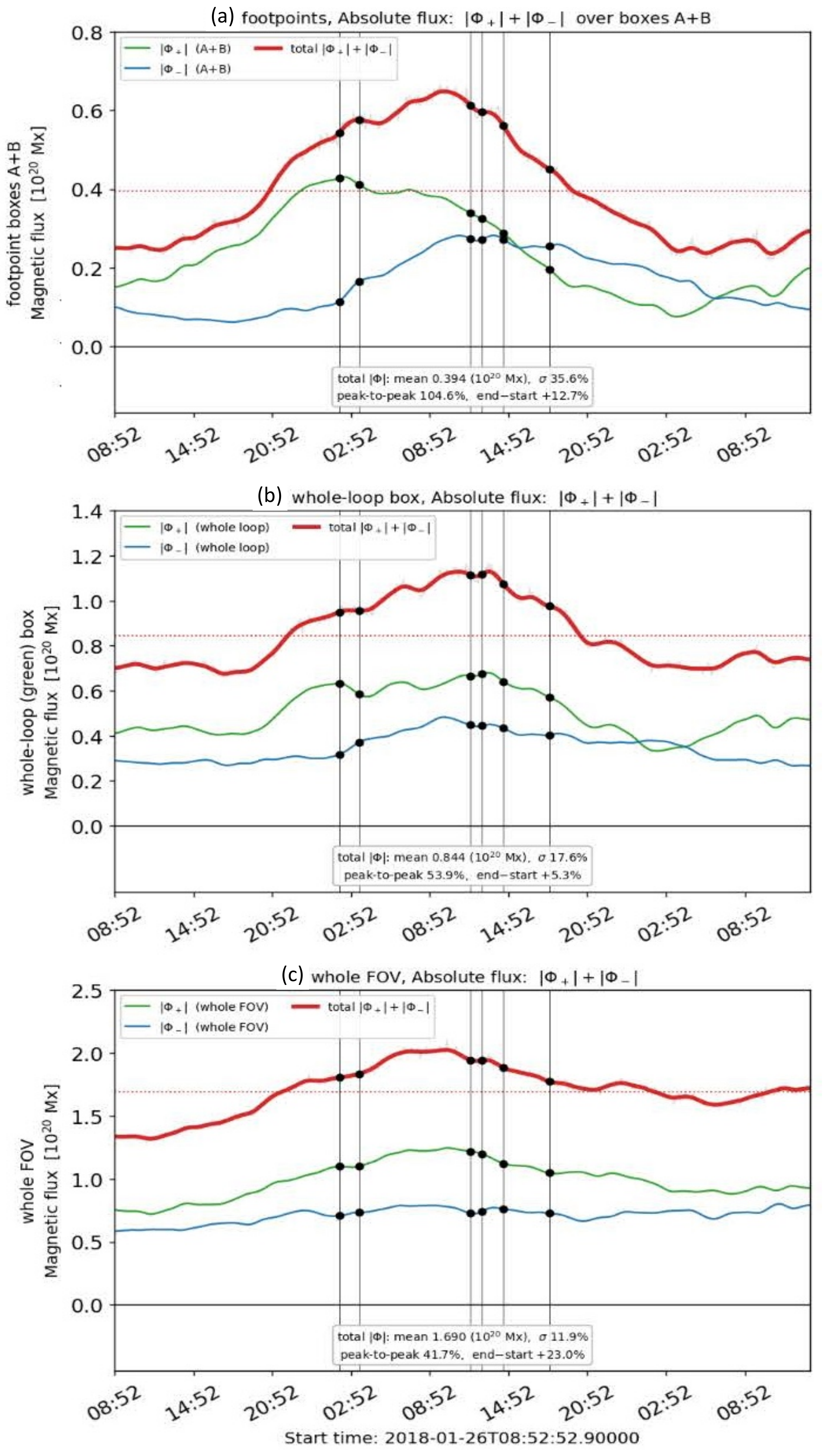}
		\caption{Absolute total magnetic flux for different regions associated with loop 3 from Table \ref{tab:1}. The panels show the absolute total magnetic flux measured at the loop feet (a), within the boxed region (b), and over the full field of view shown in the lower panels of Figure \ref{fig:1_context} (c). Since both loop feet exhibited flux convergence for this loop, we include the flux evolution for both positive and negative polarities. No threshold was applied to these flux measurements, unlike those shown in Figure \ref{fig:2_plots}.
        } 
		\label{fig:Flux_plots}
	\end{figure}

	\begin{figure*}
		\centering
		\includegraphics[trim=2cm 51cm 3cm 15cm,clip,width=\linewidth]{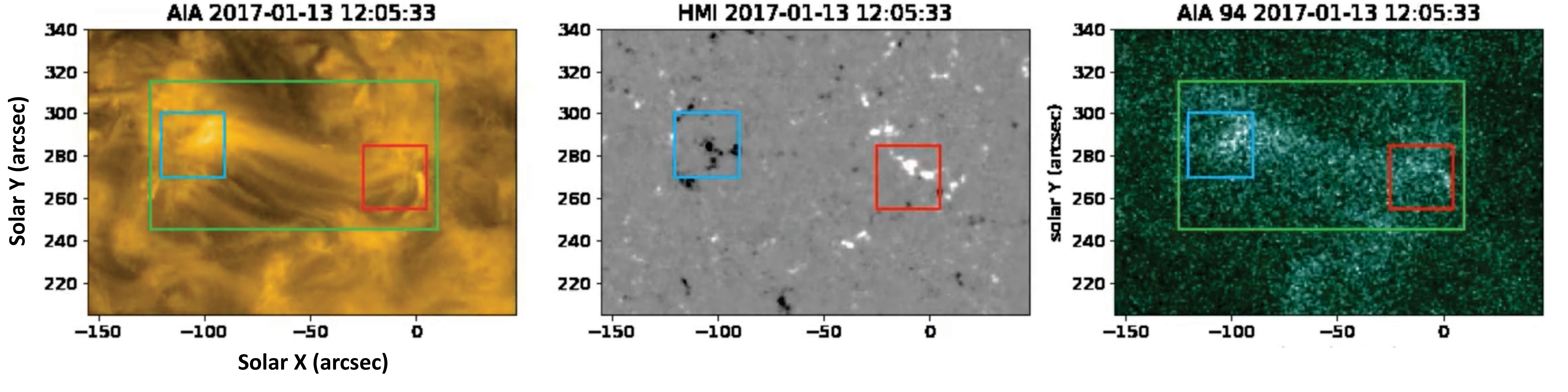}
		\caption{An example of an asymmetrically heated loop. From left to right, AIA 171 \AA, HMI LOS magnetograms, and AIA 94 \AA\ images are shown. The red box is the positive-polarity footpoint, and the green box outlines the selected area of the loop.
        In this case, the loop heating starts from the left negative-polarity magnetic flux foot (blue box), which shows convergence. This is loop 2 as listed in Table \ref{tab:1}. An animation for $\sim$24 hours at a six-minute temporal cadence is available online (Movie2\_loop\_index2.mp4). The animation runs for 12 seconds. }
		\label{fig:3_asymm_loop}
	\end{figure*}

\section{Results}	\label{sec:results}
	\subsection{Results from observations}

\subsubsection{Magnetic flux convergence leading to loop heating}
\label{results_flux_convergence_loop_heating}
    
We selected ten coronal loops (listed in Table \ref{tab:1}), as described in Section \ref{selection}, with clearly visible magnetic flux convergence at, at least, one loop footpoint. For simplicity in the analysis, we selected isolated coronal loops located in quiet-Sun regions near the disk center. Further, in quiet regions, the photospheric convergence is more distinctly isolated and traceable to its loop-body effect. Figure \ref{fig:1_context} shows an example of a loop brightening with magnetic flux convergence and its disappearance with magnetic flux divergence. This loop is shown using AIA 171 and 94 \AA\ images and HMI LOS magnetograms. It is obvious from AIA 171 \AA\ images (see panel a3) that before the loop brightens, the loop feet brighten. Although less clearly than in 171 \AA, the AIA 94 \AA\ images show peak brightness during, or very close to, the peak brightness observed in 171 \AA. All ten loops show a similar correspondence in the brightness of AIA 171 and 94 channels. Some additional examples are presented in the Appendix.

The interval from the onset of magnetic flux convergence and loop formation to the end of flux divergence and loop disappearance ranges from six to 24 hours for these ten loops, as listed in Table \ref{tab:1} under ``Sequence Duration." We note that the ``Loop Lifetime" listed in Table \ref{tab:1} corresponds to the lifetime of the principal coronal loop, which forms closely at or after the peak in magnetic flux and exhibits peak luminosity. Several individual loops appear and disappear over the course of the flux convergence process, mainly identified in AIA 171 \AA\ images. Visual inspection of the EUV movies of each event listed in Table \ref{tab:1} reveals that individual loops have lifetimes ranging from six minutes to 60 minutes. For example, the animation associated with Figure \ref{fig:1_context} (Movie1.mp4) shows an individual loop in 171 \AA\ at 14:35:33 that appears in only three frames, corresponding to a maximum lifetime of 24 minutes, given the six-minute cadence used here. The same is true for a loop most clearly visible at 12:11:33. Similarly, a loop appears at 11:17:33 in only one frame. Given the uncertainties, it could have a lifetime of 6 to 12 minutes. Unless otherwise specified as an individual loop, “coronal loop” refers throughout this paper to the principal loop associated with the peak magnetic flux at a given loop footpoint.

	 
Figure \ref{fig:2_plots} shows the AIA 171 Å (as well as 94 \AA) luminosity over the loop’s lifetime. We measured and quantified magnetic flux convergence using HMI B$_{los}$ magnetograms. We compute the sum of B$_{los}$ at each footpoint over the loop’s lifetime. However, only pixels with B$_{los}$ values above a certain threshold within the region of interest (called “Critical Magnetic Flux” by \citealt{aval18}) are included in the total sum. We describe this idea as follows.
	
\cite{aval18} introduced the idea that a critical flux threshold may need to be surpassed before a flux element contributes significantly to the heating of a coronal plume. Quantitatively, we find the critical field value by selecting the range of potential B$_{los}$ threshold values between $\pm$~200~G and $\pm$~500~G. The threshold B$_{los}$ values are calculated following the method of \cite{aval18}: 
For the formation of each loop, we selected the minimum absolute pixel-field value as our potential threshold for magnetic field strength. A correlation coefficient between the computed area-averaged above-threshold B$_{los}$ profile and the AIA 171~\AA\ luminosity profile is calculated. 
From preliminary observations, it appears that – for at least one footpoint – the B$_{los}$ time series peaks before the luminosity time series peaks, potentially because loop heating is a direct consequence of B$_{los}$ changes by convergence. Hence, flux convergence plausibly occurs before the loop heating.
Therefore, before computing correlation coefficients, we smooth both profiles with a boxcar smoothing factor of 50, and use the smoothed loop luminosity and B$_{los}$ measurements in the calculation of the correlation coefficients. We first determine the correlation coefficient for thresholds ranging from 100 to 1000~G. We define the critical magnetic field strength as the threshold value that yields the maximum correlation coefficient, i.e., the correlation coefficient closest to one. Thus, the correlation coefficient plots are used to determine the critical or threshold magnetic field strength B$_{los}$. For example, in Figure \ref{fig:2_plots} (a) and (b), we show the correlation coefficient plots for the two footpoints of a loop - in each of these two plots, a dot indicates the critical B$_{los}$. 

The panels (c) and (d) in Figure \ref{fig:2_plots} show luminosity plots of the example loop (from Figure \ref{fig:1_context}) in AIA 171 and 94 \AA. The B$_{los}$ above the threshold values have been plotted for each of the loop's foot points. While the luminosities for AIA 171 and 94 \AA\ peak almost together, the B$_{los}$ peaks at the same time in one foot (A), and precedes in the other foot (B). Most likely, the AIA 94 \AA\ signal comes from the cooler component, since it matches the AIA 171.  
Both footpoints in this example exhibit magnetic flux convergence prior to the appearance of the loop in the EUV channels. Nevertheless, the convergence of the negative-polarity flux patches shows a slightly more direct temporal correlation with the EUV luminosities in the AIA 171 and 94~\AA\ channels. 

Note, for example, the first vertical line in panel (c), which marks the onset of flux convergence for (B$_{\rm los} > 240$) G. It is evident that as the magnetic field strength increases, the luminosities in both the AIA 171 \AA\ and 94 \AA\ channels also increase. This rise observed in all three plots continues until they reach their respective peaks, nearly simultaneously. Subsequently, the magnetic field strength begins to decrease, accompanied by a corresponding decline in the luminosities.

We note that the luminosities already exhibit elevated values before the onset of flux convergence above the 240 G threshold. This is likely because weaker flux elements ($< 240$ G) had begun converging earlier, leading to an increase in the loop (or the loop feet) luminosity prior to the convergence of the stronger magnetic flux. The key point, however, is that once the convergence of the stronger magnetic flux begins, a further enhancement in luminosity is observed. Likewise, as the magnetic flux decreases, the luminosities in both channels also decline. 

More generally, within our sample, flux convergence is often more pronounced at one loop footpoint than at the other (see the additional examples in the Appendix). A particularly interesting example is Loop 4, where two successive principal loops are observed (see the corresponding movie). The first principal loop forms following the convergence of positive magnetic flux, whereas the second loop forms following the convergence of negative magnetic flux. In Table \ref{tab:1}, however, we report only the timing and characteristics of the first loop for consistency.

\subsubsection{Flux convergence vs flux emergence}

While qualitative inspection of AIA images and movies, HMI LOS magnetograms, and their corresponding quantitative plots indicates that the coronal loop intensity increases as magnetic flux converges at a footpoint, we analyze the absolute magnetic flux to determine whether any flux emergence contributes to the observed increase in magnetic flux density. For this purpose, we calculate the total absolute magnetic flux in three regions: (1) the loop footpoints, (2) the entire box encompassing the loop, and (3) the full region of interest shown in the HMI LOS magnetograms in Figure \ref{fig:1_context}. The resulting flux evolution is presented in Figure \ref{fig:Flux_plots}. In all three cases, we observe an increase in the measured absolute magnetic flux, with the increase being less pronounced for larger fields of view.

    We suggest that the increase in total magnetic flux does not necessarily indicate the emergence of new magnetic flux. Instead, it may reflect the increasing detectability of magnetic flux as convergence enhances the magnetic flux density within the region. This interpretation is also supported by the absence of visible signatures of flux emergence.

    In an idealized situation, in the absence of flux emergence, submergence, or transport across the boundaries, the total magnetic flux would be expected to remain conserved. However, the measured unsigned flux can vary with time because of the finite sensitivity of the instrument. For HMI, the noise level is approximately $\pm 7$ G \citep{couv16}. Initially weak magnetic flux concentrations may therefore remain below the detection threshold. As these concentrations converge and coalesce, their field strengths can increase above the detection threshold, leading to an apparent increase in the measured unsigned flux. Conversely, as magnetic flux concentrations diverge and their field strengths decrease below the detection threshold, the measured unsigned flux can decrease. Thus, temporal variations in the measured unsigned flux do not necessarily imply the creation or loss of magnetic flux, but can instead partly reflect the finite sensitivity of HMI and the resulting changes in the detectability of weak-field regions. We therefore interpret the observed increase in unsigned flux primarily as a consequence of flux convergence and the associated increase in field strength, rather than as evidence of flux emergence. We also note that some magnetic flux elements may cross the boundaries of the box, entering or exiting the box during the sequence.

\subsubsection{Asymmetric loop heating}
 
The loop heating in the example presented in Figure \ref{fig:1_context} is symmetric, with both footpoints brightening nearly simultaneously, followed by the rapid appearance of the loop. However, as listed in Table~\ref{tab:1}, several loops exhibit asymmetric heating, characterized by a clear propagation of brightening from one footpoint to the other (see Figure~\ref{fig:3_asymm_loop} for an example; additional cases are presented in the Appendix). Notably, in these asymmetric events, the footpoint that brightens first is also the one where magnetic flux convergence is observed to occur first, and in some cases, the only footpoint at which flux convergence is detected.

As noted for the positive-polarity footpoint of the example presented here, the magnetic field time series often peaks before the EUV luminosity (Figure~\ref{fig:2_plots}d). In all cases, the EUV luminosity increases nearly simultaneously with, or after, an increase in magnetic flux at least at one of the footpoints.

Two additional observational examples presented in the Appendix exhibit similar behavior: at one footpoint (the positive-polarity footpoint in both loops), the magnetic field and luminosity peak nearly simultaneously, or the magnetic field peaks slightly earlier, whereas at the other (negative-polarity) footpoint, the magnetic field peaks substantially later than the luminosity. Cases in which the magnetic field peaks before, or nearly simultaneously, with the luminosity at one or both loop footpoints are more intuitive, as they are consistent with loop heating driven by magnetic flux convergence. We note that, in every loop in our sample, at least one footpoint exhibits a peak in magnetic field strength either shortly before or nearly simultaneously with the luminosity peak. In cases of asymmetric loop heating, the heating onset occurs at that footpoint.

Again, while in most loops both footpoints show magnetic flux convergence, it is often more pronounced at one footpoint than at the other. Table~\ref{tab:1} lists the magnetic flux values at that loop footpoint corresponding to the peak magnetic field strength. For the ten loops, the magnetic flux values, during the peak flux, range from $1\times10^{19}$ to $5.1\times10^{19}$~Mx.

All ten examples listed in Table~\ref{tab:1} exhibit a similar correspondence between EUV luminosity and magnetic field as described above. To further explore their origin, we searched for an analogous correspondence in a Bifrost MHD simulation of a network region containing numerous coronal loops, as described in the next section.

\subsection{Simulation results}
\label{ref:subsection:simulation_results}

\subsubsection{Visualizing and selecting the loops}
\label{ref:subsubection_loops_vis}
\begin{figure*}
    \centering
    \includegraphics[width=\linewidth]{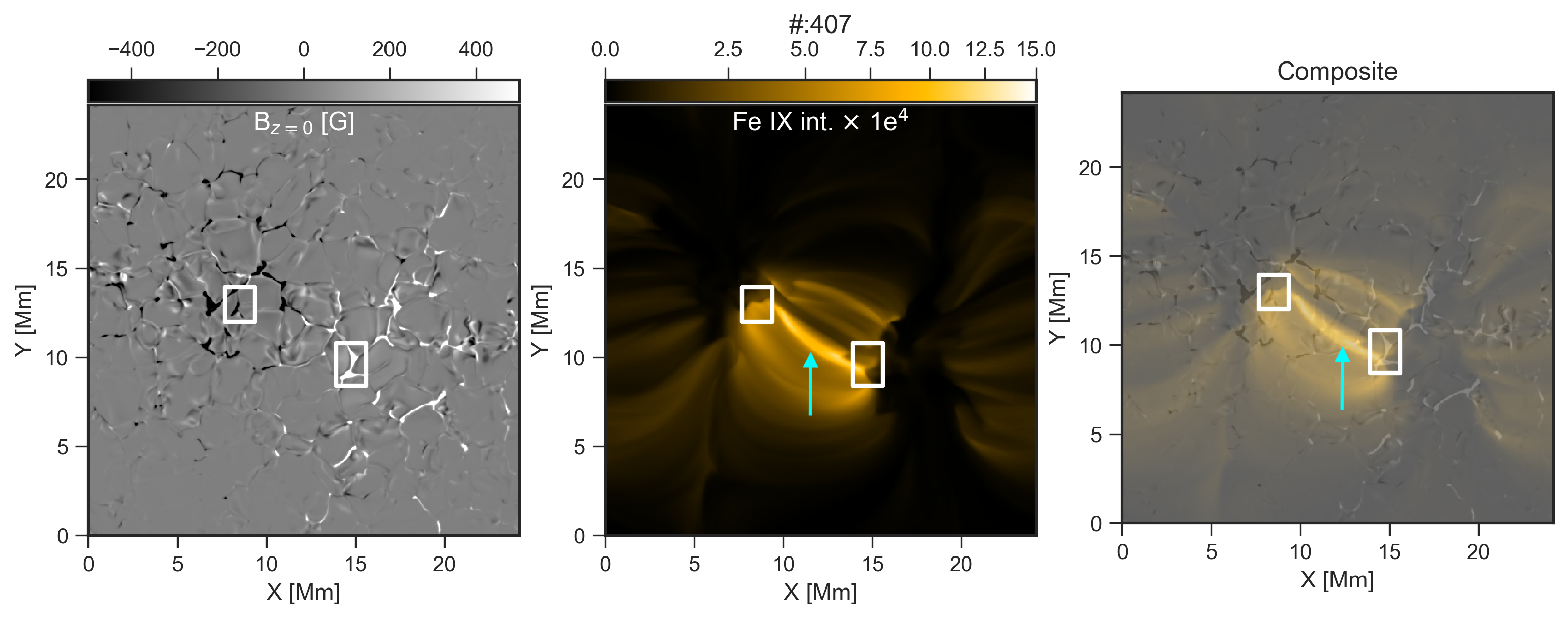}
    \caption{Visualizing the loop system of interest and its photospheric vertical magnetic field configuration. From left to right, the panels show the signed vertical component of the magnetic field strength ($B$) at $z=0$, the \ion{Fe}{ix} intensity image, and a composite of the two. The cyan arrows point to the loop system analyzed in this paper, and the rectangular boxes mark the footpoint regions. The number \#407 at the top is the snapshot number of the simulation, that is 4070~s after the start of the simulation (\telap) and 1050~s after switching on the non-equilibrium ionization of hydrogen. An animation (Movie5.mp4) covering the entire temporal evolution of the loop between snapshots 323 (\telap=3230~s) and 480 (\telap=4800~s) is available online. The animation runs for seven seconds.} 
    \label{fig:sim_example_loop}
\end{figure*}

We follow an approach similar to the observations described above to identify a loop in the enhanced network simulation qualitatively. We visualize the variation of synthetic  \ion{Fe}{ix} intensity (analogous to the brightness in the SDO/AIA~171 \AA\ observations) along with B$_{\mathrm{z}}$ at the surface to select a loop connecting the two opposite-polarity patches of the enhanced network at the loop's feet (Figure~\ref{fig:sim_example_loop}). The indicative footpoints of the loop show several episodes of flux convergence and divergence, as evidenced by the animation of B$_{\mathrm{z}}$ shown in the figure.

\begin{figure*}
    \centering
    \includegraphics[trim={0cm 0.1cm 0.3cm 0.3cm},clip=True,width=\linewidth]{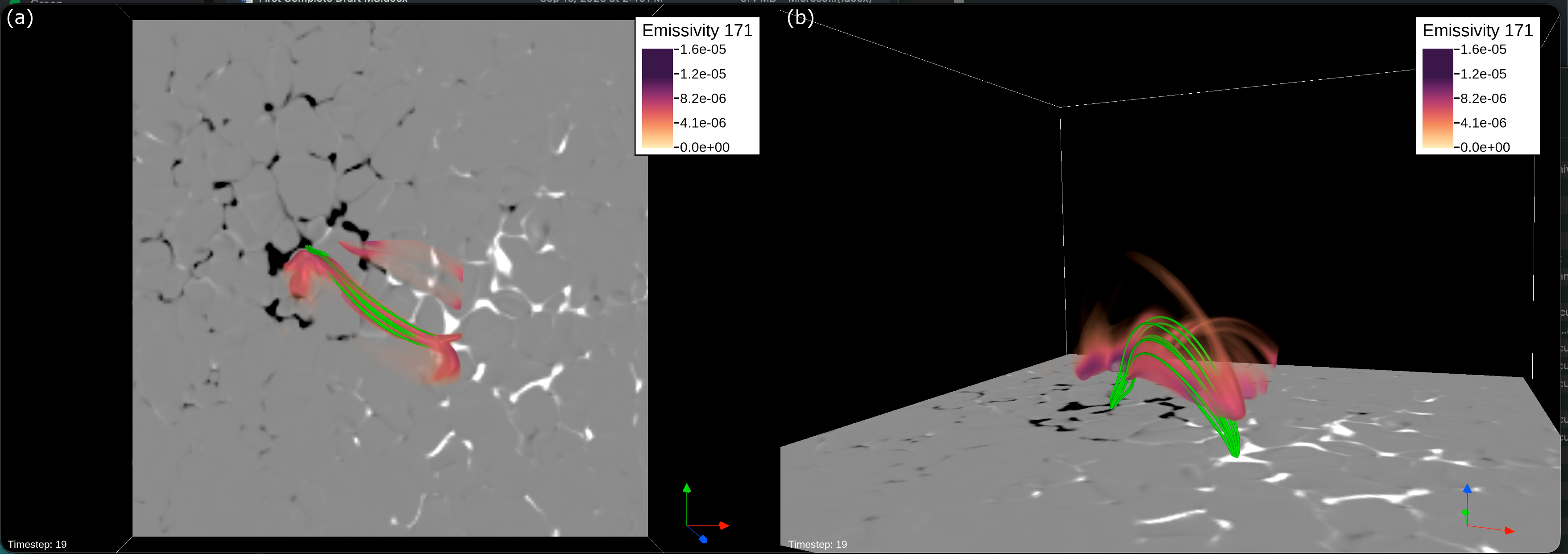} 
    \caption{3D rendering of the loop's emissivity and its corresponding bundle of selected field lines at snapshot number 419 (\telap = 4190~s). Panels~(a) and (b) show the top and the side view of the signed magnetic field strength ($B$) at $z=0$, the emissivity of \ion{Fe}{ix}, and the 3D configuration of the bundle of ten selected field lines covering the volume of the loop. An animation (Movie6.mp4) covering the sequence of every 4 snapshots (at a cadence of 40~s ), ranging from 323 (\telap=3230~s) to 480 (\telap=4800~s) is available online. The animation runs for four seconds. }  
    \label{fig:sim_loop_3D_lines}
\end{figure*}

Following this, the corresponding \ion{Fe}{ix} emissivity of the loop is visualized in 3D using \verb|Vapor| \citep{li19}. A bundle of 10 magnetic field lines, roughly covering the loop's emissivity volume, is selected for further visualization and analysis. The field lines are chosen by biasing with the emissivity strength. We used these field lines to integrate the emissivity over the loop, henceforth referred to as the volume-integrated emissivity (e.g., black lines in panels (g) and (h) of Figure~\ref{fig:sim_div}). Figure~\ref{fig:sim_loop_3D_lines} and its animation show the loop's 3D temporal evolution from snapshots 323 to 480 with a cadence of 10~s\footnote{Refer to ITN 33 of \url{https://iris.lmsal.com/documents.html}}, which is about 27~min in duration. We note that this typical lifetime of individual coronal loops is also found in our observations. Like in the observations, the standard contrast stretching method ($\gamma$ correction) is applied to enhance the visibility of this particular loop at each time step and suppress others in the close vicinity.

The animation associated with Figures~\ref{fig:sim_example_loop} and \ref{fig:sim_loop_3D_lines} clearly shows the presence of unipolar (opposite) polarity patches at the two footpoints of the loop system. Flux convergence is also seen at both footpoints and appears to be closely associated with the enhancement of the volume-integrated \ion{Fe}{ix} emissivity. The convergence causes the magnetic field lines to be impacted by photospheric motions, eventually leading to continuous and sporadic reconnection and reconfiguration of the magnetic field topology. Figure~\ref{fig:sim_loop_3D_lines} shows the overlap between the selected field lines and the volume-integrated emissivity. In this paper, we study in detail the relationship between flux convergence and the subsequent appearance, energy propagation, and heating of the selected \FeIX\ loop.

\subsubsection{Relationship between flux convergence and vorticity with loop formation}
\label{ref:subsubsection_convergence_vorticity}

\begin{figure*}[ht!]
    \centering
        \includegraphics[width=0.91\linewidth]{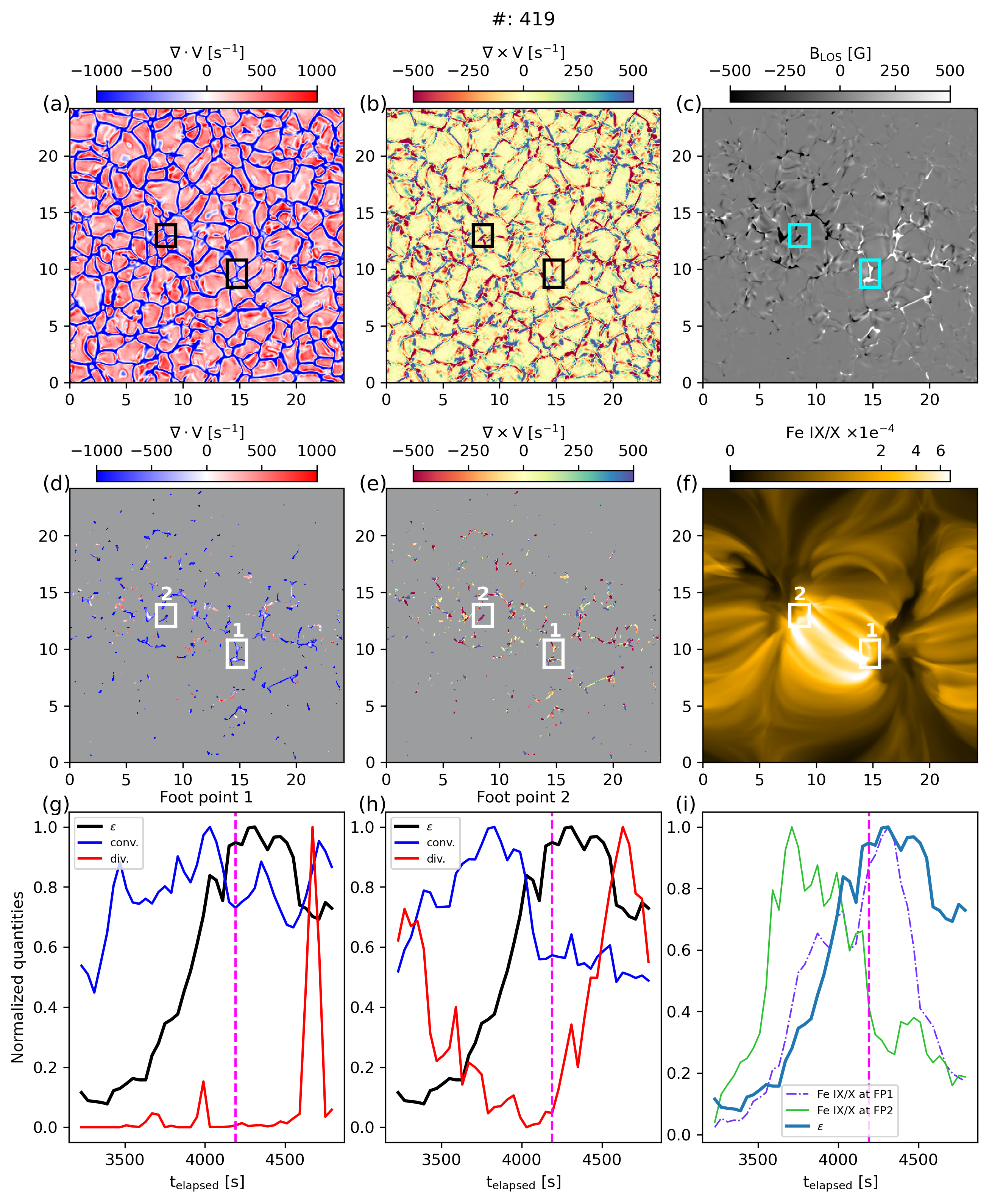}
    \caption{Relationship between flux convergence (divergence), shear, and the volume-integrated emissivity of the coronal loop system at snapshot 419 (\telap=4190~s). Panels~(a) and (b) show the divergence and the curl (shear) of the horizontal velocity vectors at heights where $\tau_{500}=1$ (see text for more details). Panel~(c) shows the line-of-sight magnetic field strength at $z=0$. Panels~(d) and (e) show divergence and the curl of the horizontal velocity vectors, but only for pixels above the critical magnetic field threshold (200~G). Panel~(f) shows the synthetic \ion{Fe}{ix} LOS luminosity. Panels~(g) and (h) show the temporal evolution of the magnitude of the scaled (to their respective maximum) convergence and divergence at footpoints~1 and 2, and the volume-integrated emissivity  ($\epsilon$) of the loop. Panel~(i) shows the comparison of $\epsilon$ against the \ion{Fe}{ix} luminosities at the two footpoints. The vertical dashed magenta line in the bottom row indicates the \telap. An animation (Movie7.mp4) covering the entire temporal evolution of the loop between snapshots 323 (\telap=3230~s) and 480 (\telap=4800~s) is available online. The animation runs for four seconds. }
    \label{fig:sim_div}
\end{figure*}

We compute 2D maps of divergence and the vorticity (Figure~\ref{fig:sim_div}~a and b) from the horizontal velocities of the simulation at heights where $\tau_{500}=1$ to study the effect of flux convergence and shear (if any) on the brightness of the \ion{Fe}{ix} loops. Although we do not perform this analysis in observations because of the large noise in the transverse field measurements of HMI magnetograms for QRs, vortices may nevertheless play an important role in chromospheric and coronal loop heating \citep{bran88,zirk93vortices,yada21,breu23}.
The $\tau_{500}=1$ heights are chosen to minimize the impact of $p$-mode oscillations, which are present in the whole simulation domain \citep{carl16}. The oscillations can vary the average height at optical depth unity in the snapshots by approximately 60~km.

Furthermore, we performed a cross-correlation analysis using the LOS magnetic flux density (Figure~\ref{fig:sim_div}~c) and the \FeIX\ luminosity (Figure~\ref{fig:sim_div}~f) maps at the two footpoints derived from the simulation, and found that a critical unsigned magnetic flux density of $\approx$200~G is required to achieve the highest correlation between the two quantities. The details of this conclusion can be found in   Figure~\ref{fig_appendix:critical_field_sim} of Appendix~\ref{appendix_sim}. The idea of the critical magnetic field stems from the observations described in the paper, where it is hypothesized that a given magnetic flux element needs to be beyond a certain threshold to contribute significantly towards the enhanced heating and brightening coronal loops \citep[as suggested in][]{aval18}. However, unlike in the observations, we do not select the entire loop because it is difficult to isolate individual loops in the synthetic luminosity maps and the model allows to go into more detail and isolate the 3D rendered loop from the rest along the LOS. Therefore, we select only pixels from the divergence and vorticity maps that have an absolute magnetic flux density greater than 200~G for further analysis, as described below. We emphasize that the selection of 200 G is consistent with that of observations. Maps showing the selected measured pixels are outlined in panels~(d) and (e) of Figure~\ref{fig:sim_div}. 

The bottom row, i.e., panels~(g)--(i) of Figure~\ref{fig:sim_div} shows the temporal evolution of the normalized (to the respective maximum) divergence, convergence, the volume-integrated \FeIX\ emissivity, and the \FeIX\ luminosity, and attempts to explore the relationship among them. Panels~(g) and (h) clearly show an increase in the flux convergence before the volume-integrated emissivity peak. Furthermore, we also see negligible divergence in footpoint~1 (panel~g) and a gradual decrease in the divergence in footpoint~2 (panel~h). The combination of the two suggests a relationship between flux convergence and subsequent loop brightness, as hypothesized in this paper from the observations. However, it is difficult to understand whether one or both footpoints are equally important contributors to the loop heating. Figure~\ref{fig:sim_div}~panel~(i) shows the importance of considering 3D volume-integrated emissivity in capturing the loop brightening instead of selecting the footpoints. Selecting only the squares at the loop footpoints can yield different intensities than other parts of the loop and can therefore fail to capture the entire scenario. E.g., the \FeIX\ brightness at footpoint 1 shows a closer relationship with the volume-integrated emissivity compared to footpoint 2. However, the loop, as shown in 3D in Figure~\ref{fig:sim_loop_3D_lines}, encompasses both footpoints along with its apex. 

\begin{figure*}
    \centering
    \includegraphics[width=0.8\linewidth]{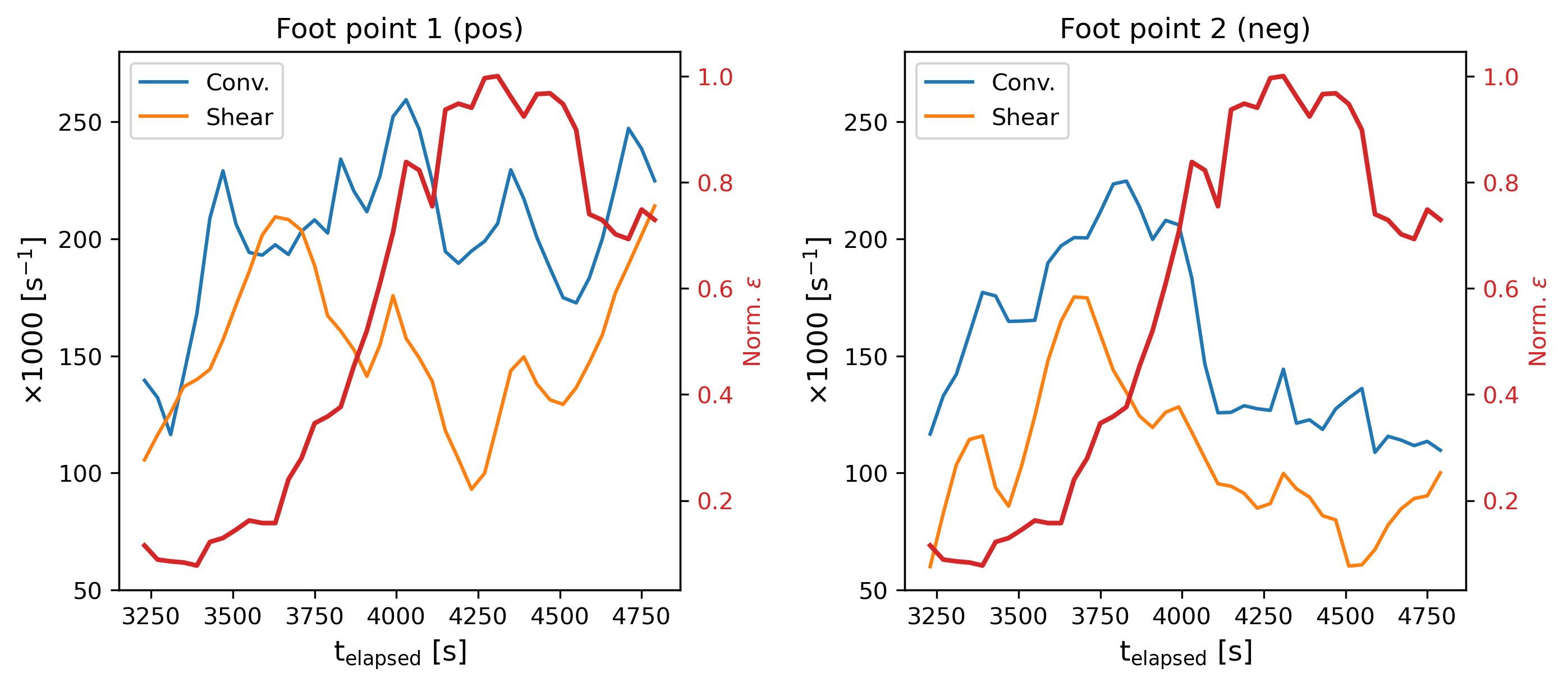}
    \caption{A comparison of the temporal evolution of the magnitude of the convergence, shear, and the volume-integrated emissivity ($\epsilon$, scaled to its maximum) at footpoint~1 (left panel) and footpoint~2 (right panel) of the loop system being studied. The temporal duration covers 1600~s from 3250~s after the start of the simulation.} 
    \label{fig:sim_vort_shear}
\end{figure*}

The impact of shear $\left| \nabla\times V \right|$ and its possible link with the loop's volume-integrated emissivity at the two footpoints is also shown in Figure~\ref{fig:sim_vort_shear} as a function of time. Although both footpoints show an increase in the shear and flux convergence, reaching their respective peaks before the volume-integrated emissivity peak of the loop, it is clear that the two quantities are stronger (see their magnitudes) in footpoint~1 (left panel) compared to footpoint~2 (right panel). As we will address below, it is also footpoint~1 that leads to more Joule heating.  
Flux convergence and shear both increase and decrease nearly simultaneously, although with a slight time lag, as the decrease in shear precedes that of convergence. This suggests that the initial flux convergence may trigger the development of shear. However, disentangling the respective roles of shear and convergence is extremely challenging. It is possible that flux convergence induces shear in the magnetic flux elements, which in turn drives the heating. Interestingly, the flux convergence continues to rise in footpoint 1, peaking around 4000~s, whereas at the other footpoint, the peak is advanced by 250~s (reaching a maximum around 3750~s). To further understand the relative importance of the two footpoints, we compute and analyze the energy transfer in the form of Poynting flux as described in the following section.

\subsubsection{Poynting flux at the two footpoints of the loop}
\label{ref:subsubsection_poynting}

The heating of the loops is tied to the magnetic energy flux that is carried to and dissipated in the upper atmosphere. This energy flux comes from the photosphere in the form of Poynting flux,  $S=(\bold{E}\times\bold{B})/\mu_{0}$. The vertical component of this flux ($S_{z}$) can mathematically be decomposed into two main parts \citep[e.g.,][]{manchester2004,hans15}. The first is the emerging Poynting flux due to the transport of the horizontal field by vertical motions,
\begin{equation}
\label{equation_1}
    S_{z,emer}=\frac{v_{z}}{\mu_{0}}(B_{x}^2+B_{y}^2)
\end{equation}
and the second is due to the shear or the work done by the horizontal motions on the vertical field, 
\begin{equation}
\label{equation_2}
    S_{z,shear} =-\frac{B_{z}}{\mu_{0}}(v_{x}B_{x}+v_{y}B_{y})
\end{equation}

There is a third part due to the finite magnetic diffusivity that is often negligible compared to the two parts above \citep{2022A&A...665A.118F}. While the energy is carried by both the vertical (emerging) and horizontal (shearing) components, studies \citep[e.g.,][]{2022ApJ...927..146S} have shown that the total Poynting flux near the photosphere is primarily horizontal and concentrates over small regions along intergranular lanes. Moreover, the correlation between flux convergence (and shear) in the two footpoints with the loop's volume-integrated emissivity discussed in subsection~\ref{ref:subsubsection_convergence_vorticity} further motivates the choice of considering the Poynting flux through only equation~\ref{equation_2} in all subsequent analysis.

\begin{figure*}[ht!]
    \centering
    \includegraphics[trim={0cm 0cm 0cm 0cm},clip=True,width=\linewidth]{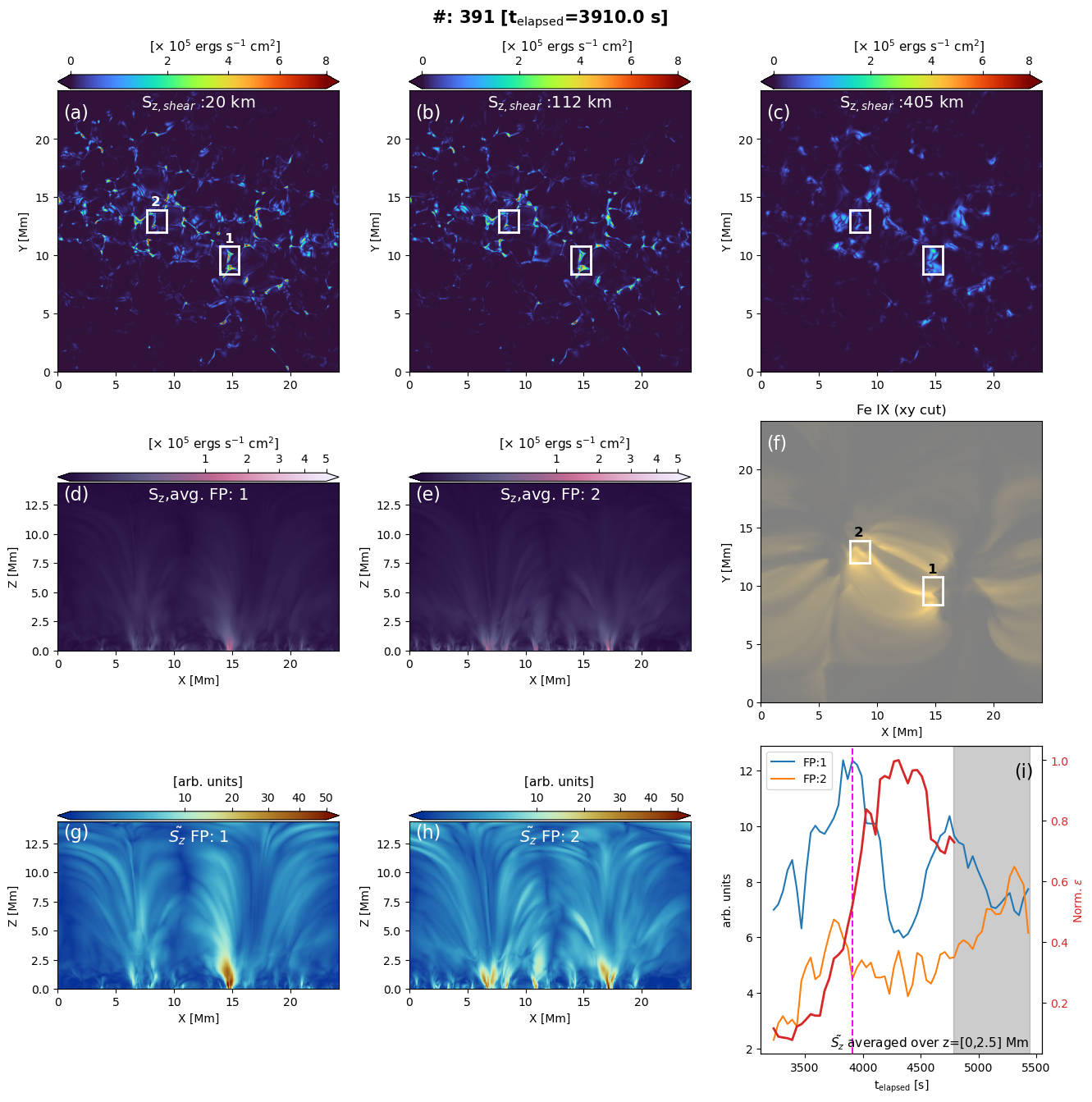}   
    \caption{Vertical Poynting flux ($S$) associated with the loop system and its relationship with volume-integrated emissivity ($\epsilon$) at \#391 ($t=3910$~s). Panels~(a)--(c) show the vertical component of the $S$ due to the shearing motions (c.f.~\ref{equation_2}) at heights $20$, $112$, and $405$~km from $z=0$ surface. The rectangular boxes in the three panels show the footpoints (1 and 2) of the loop under investigation. Panels~(d) and (e) show the $xz$ maps of the Poynting flux averaged along the $y$ direction of boxes~1 and 2, respectively. Panel~(f) shows the \ion{Fe}{ix} luminosity with the opacity intentionally set to lower values to allow better focus on the loop. Panels~(g) and (h) show $xz$ maps of the dimensionless quantity ($\Tilde{S_{z}}$, c.f. eq.~\ref{equation_2}) for the two footpoint regions 1 and 2 in the same format as panels~(d) and (e). Panel~(i) shows the temporal evolution of $\Tilde{S_{z}}$ at the two footpoints averaged between $z=[0,2.5]$~Mm along with the volume integrated $\epsilon$ (in red). The shaded gray region indicates the time interval where the $\epsilon$ of the loop under investigation fades, and a new loop (bundles) starts appearing beyond $t_{\mathrm{elapsed}}=4800$~s. The vertical dashed magenta line in the bottom row indicates the \telap. An animation (Movie8.mp4) covering the duration between snapshots 323 (\telap=3230~s) and 539 (\telap=5390~s) is available online. The animation runs for six seconds.}
    \label{fig:poynting_flux}
\end{figure*}

\begin{figure}[htp]
    \centering
    \includegraphics[width=\linewidth]{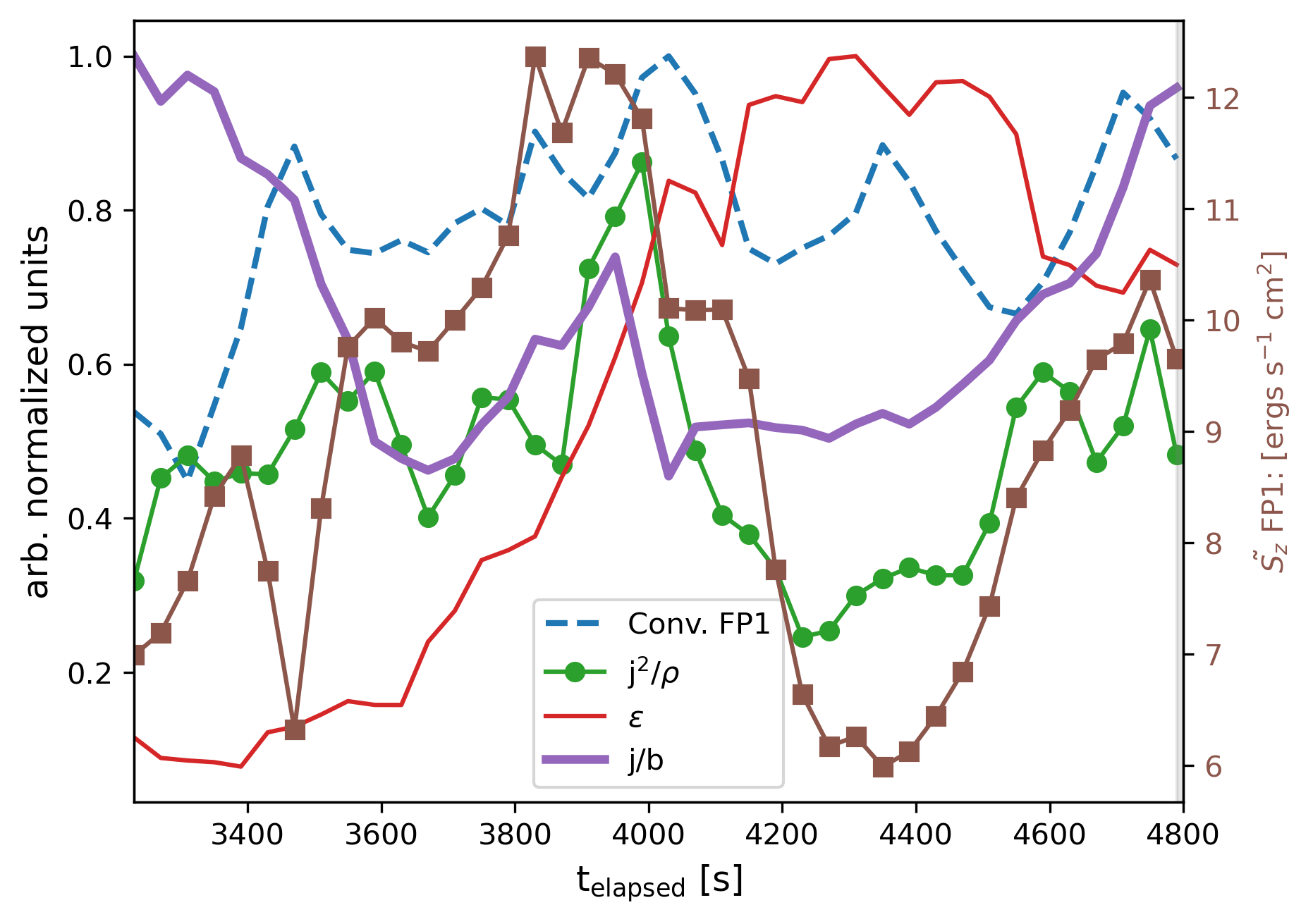}
    \caption{Deciphering the physics of heating of the coronal loop. The figure shows the temporal evolution of the dimensionless $\Tilde{S_{z}}$ (brown) along with the flux convergence (blue), Joule heating per particle ($\frac{j^2}{\rho}$, green), inverse characteristic length ($\frac{j}{b}$) (scaled to their respective maximum, purple) at footpoint~1. The scaled volume-integrated $\epsilon$ is also shown for comparison (red). The quantities, $\frac{j^2}{\rho}$ and $\frac{j}{b}$, are averaged between $z=[0,2.5]$~Mm in the same way as $\Tilde{S_{z}}$. The overall temporal duration spans from $t=3230$~s to $t=4800$~s after the start of the simulation. }
    \label{fig:sim_physics}
\end{figure}

Figure~\ref{fig:poynting_flux} and its associated animation show the $xy$ slices of $S_{z,shear}$ at three geometrical heights ($z$, top row), and $xz$ slices of averaged shear for a range of $y$ values $S_{z,avg}$ (covering the two footpoints marked as solid boxes 1 and 2) in panels~(d) and (e). The animation clearly shows the enhancement in $S_{z,shear}$ within the two solid boxes in the top row across all the geometrical heights, and a close look also suggests that the enhancement associated with flux convergence is stronger in box~1 covering footpoint~1. The enhancement in $S_{z,shear}$ in the region bounded by the two footpoints is also evident in $xz$ maps ($S_{z,avg}$ panels~(d) and e). To further accentuate the impact of the Poynting flux, we compute a dimensionless quantity, 
\begin{equation}
    \Tilde{S_{z}}=S_{z}/\overline{S_{z,xy}}
\end{equation}
where $\overline{S_{z,xy}}$ represents the mean $S_{z}$ over the $xy$ plane at a given geometrical height $z$, and plot the corresponding $xz$-slices in panels~(g) and (h) of Figure~\ref{fig:poynting_flux}. The two panels and their temporal evolution clearly show strong $\Tilde{S_z}$ at the two footpoints, with footpoint~1 stronger than the other. Panel~(i) shows the temporal evolution of the vertically averaged $\Tilde{S_z}$ between $z=[0,2.5]$~Mm, overplotted with the volume-integrated emissivity. Judging by their respective evolution, it is clear that the peak $\Tilde{S_z}$ of footpoint~1 is roughly twice that of footpoint~2, and occurs closer to the peak of the emissivity light curve. This suggests that footpoint~1 is the dominant contributor of energy to the overlying simulated coronal loop system. Therefore, in the following section, we attempt to understand the physics of the heating of the loop by focusing solely on footpoint~1.

\subsubsection{Dynamics at loop footpoint~1}
\label{ref:subsubsection_physics_footpoint1}

Figure~\ref{fig:sim_physics} shows the temporal evolution of the flux convergence, $\Tilde{S_z}$, Joule dissipation per particle (also called current density squared per particle, $j^2/\rho$), and the inverse characteristic length of the magnetic field variation \citep[j/b,][]{nobr16,2024NatAs...8..697B} at footpoint~1 compared against the overall volume-integrated emissivity ($\epsilon$). The inverse characteristic length ($j/b$) is a quantity that permits visualization of any abrupt changes in the magnetic field and is linked with current sheets formed due to magnetic reconnection \citep{prie14}. It is interesting to note that overall peaks of $j^2/\rho$, $j/b$, and $\Tilde{S_z}$ occur around the same time (\telap $\approx$3950~s), which correlate well with the peak of the flux convergence. The $\epsilon$ peak is delayed by about 400~s compared to the other quantities. 
We note that the convergence plot is not ideally clean, in the sense that it does not show a systematic rise and fall with time. This may result from magnetic flux entering and leaving the selected region, and could reflect the difficulty of isolating all relevant flux patches in the simulation. Nonetheless, the overall temporal evolution presented here with other parameters  suggests a scenario for the heating and subsequent brightening of quiescent coronal loops, as discussed in Section~\ref{Sect:discussion}. We also find that the total Joule heating and currents are larger at footpoint~1, suggesting that footpoint~1 may play a major role in the formation of this loop.

\subsubsection{Is there any magnetic flux emergence that could contribute to the flux increase?}
\label{ref:abs_flux_plots_footpoint1}

\begin{figure}[htp]
    \centering
    \includegraphics[width=\linewidth]{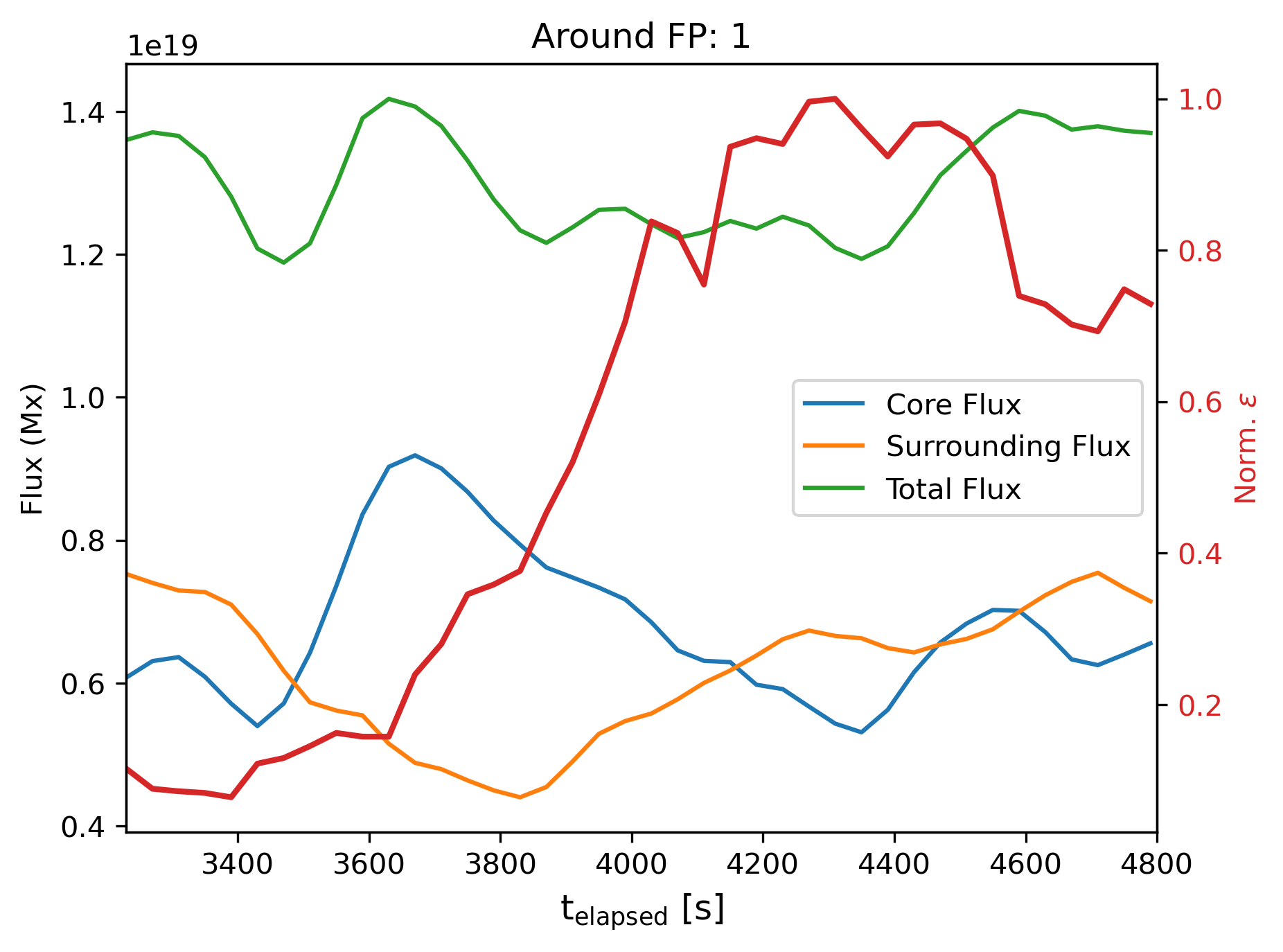} 
    \caption{Magnetic flux plots for footpoint 1 are shown to assess the contribution of any flux emergence. The temporal evolution along the X-axis is the same as in Figure \ref{fig:sim_physics}. The scaled volume-integrated $\epsilon$ is also overplotted in red.}
    \label{fig:sim_flux_plots}
\end{figure}

Here, we investigate whether magnetic flux emergence at Footpoint 1, rather than flux convergence, contributes to the observed increase in flux density. To distinguish between these possibilities, we analyze the local flux budget by tracking two nested regions: the tight footpoint core and a wider surrounding region extending 15 pixels on either side of the core. The resulting flux budgets are shown in Figure \ref{fig:sim_flux_plots}.

During the period of t = 3400 to 3700\text{ s}, the core flux (blue) increases rapidly from \(\sim 0.55 \times 10^{19}\) to a peak of \(\sim 0.92 \times 10^{19} \text{ Mx}\). At the same time interval, the surrounding flux (orange) decreases from \(\sim 0.72 \times 10^{19}\) to a minimum of \(\sim 0.52 \times 10^{19} \text{ Mx}\).

We note that the total flux (green) remains approximately stable and bounded during this main migration phase, suggesting that the observed increase in core flux is not due to the emergence of new magnetic flux through the photosphere. Instead, the simultaneous increase in core flux and decrease in surrounding flux indicate that existing magnetic elements are being transported from the surrounding region into the core and concentrated there.

As the flux convergence reaches its peak at t $\approx$ 3600 s, the normalized emissivity begins to increase rapidly, increasing from 0.2 to more than 0.9. This temporal correspondence indicates that the flux-convergence phase precedes the rapid formation of the bright loop, suggesting a close relationship between the concentration of magnetic flux and loop formation.

Overall, this analysis shows that the increase in core flux is primarily driven by local flux convergence rather than by the emergence of new magnetic flux.

\section{Discussion}
\label{Sect:discussion}

		\begin{figure*}[htp]
				\centering
                \includegraphics[trim=0cm 0cm 0cm 0cm,clip,width=\linewidth]{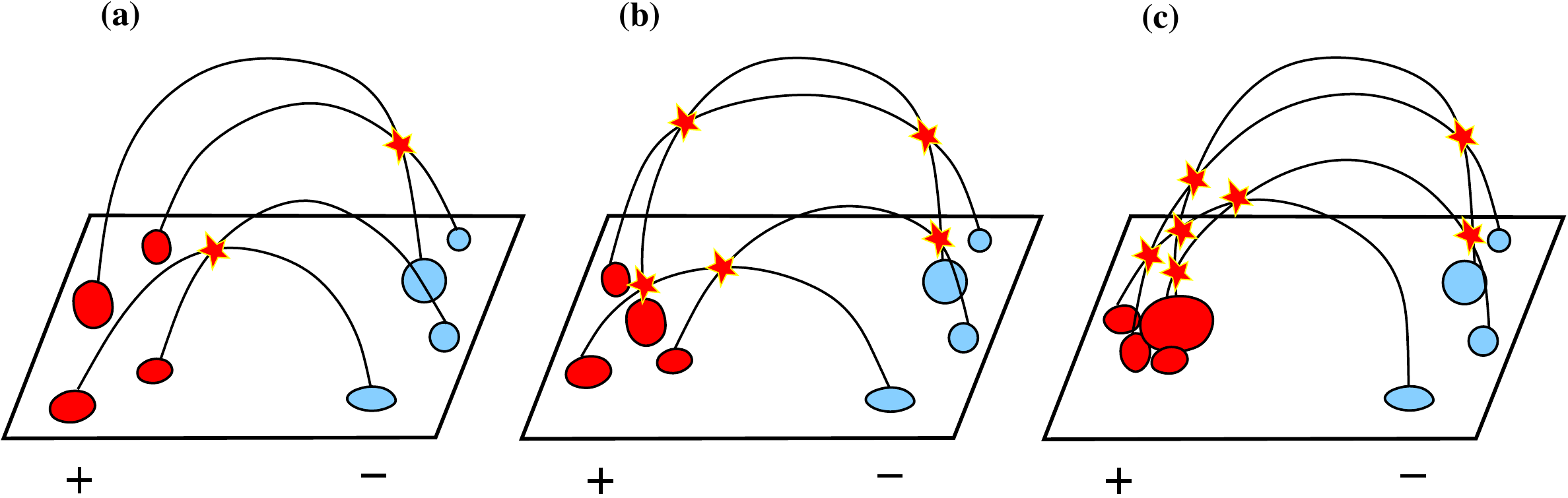}
				\caption {Drawings depicting the loop heating process by photospheric magnetic flux convergence. Panels (a), (b), and (c) depict three stages during the convergence and loop heating process. Flux convergence in one foot (or both feet) of a loop system, or a sequence of loops, can lead to an increased number density of nanoflares, thus increased heating in the loop system. This simple diagram shows most convergence in the positive polarity flux with insignificant change in the negative polarity flux. Asymmetric heating would occur, with the heating intensity propagating from the positive- to the negative-polarity side of the loop. The positive magnetic polarity is represented by red-filled circles, and the negative magnetic polarity is represented by blue-filled circles. Asterisks indicate magnetic reconnection, the nanoflare process. 
              }
				\label{fig:drawing}
			\end{figure*}

From observations of ten sets of coronal loops, including measurements of their EUV luminosity and loop-footpoint magnetic field throughout their lifetimes, and from a detailed analysis of a synthetic loop produced by a radiative Bifrost MHD simulation, we find that magnetic flux convergence and subsequent divergence at the loop footpoints are correlated with the loop brightening in EUV emission and their subsequent disappearance, respectively. Thus, the combined analysis of the observed and synthesized loops derived from the 3D radiative-MHD simulation provides important insight into the physical processes governing the formation, evolution, and disappearance of coronal loops.

We selected isolated loop systems that were sufficiently separated from neighboring coronal structures to minimize the influence of interactions with adjacent structures on their evolution. External triggering can induce episodic heating events in coronal loops, as demonstrated by the observations of \cite{tiw14} and the simulations of \cite{arch14}. By restricting our analysis to isolated loops, we can better isolate and assess the role of magnetic flux convergence and divergence at the loop footpoints in governing loop evolution.

A critical magnetic field strength in the range of approximately 200–500 G at at least one loop footpoint appears necessary for heating and consequent visibility of the observed loops in the AIA 171 \AA\ channel. However, not all instances of magnetic flux convergence on the Sun necessarily lead to loop formation, analogous to the fact that not all comparable flux-convergence events result in plume formation, as discussed by \cite{wang16} and \cite{aval18}. It is noteworthy that plume formation on the Sun has not been observed without magnetic flux convergence at its photospheric footpoint \citep{aval18}, but see \cite{weit25} for sunspot plumes. While this requirement may not universally apply to the heating of all coronal loops, our detailed analysis of the ten loops presented here, combined with extensive visual inspections of solar QRs using JHelioviewer \citep{mull17_JHelioviewer}, did not identify any cases in which a loop formed without evidence of magnetic flux convergence at at least one footpoint. In particular, the large time lags between magnetic flux and luminosity in some of our observations may indicate that, as in the case of coronal plumes, flux convergence is necessary but not sufficient for coronal loop heating. This suggests that an additional physical factor must be incorporated to fully understand the heating mechanisms of these loops.
A statistical analysis of loops derived from simulations should be carried out in future studies to establish a more robust conclusion. 

We find that there is usually a delay between the peaks of the loop-luminosity and loop-foot magnetic flux evolution.
For the loops in which foot-flux convergence led the brightening, the magnetic flux reached its peak from six minutes to a couple of hours before the peak of the main loop brightening. We note that this delay is typically less than 30 minutes. In cases where the time lag between the peak of flux convergence and the peak in luminosity exceeds 30 minutes, it may indicate that the flux convergence triggers a sequence of loop-heating events, resulting in the formation of multiple loops in succession \citep[i.e., a nanoflare storm:][]{klim09,hood16,barn16}. Alternatively, this may indicate that the flux concentration could facilitate the development of other small-scale dynamic phenomena at later times, contributing to coronal heating, such as spicules \citep[e.g.,] []{ster00,depo07_spicules,mart17Sci,martinez2018,bose23,bose2025,pane26_dkist}, jetlets \citep[e.g.][]{raouafi14,pane18a,pane19}, and microflashes \citep{pane26_microflashes}, among others, which are observed in quiet-Sun network regions and at plume bases.
Nonetheless, that footpoint flux convergence peaks simultaneously or before the loop reaches its strongest intensities suggests that convergence-driven heating is fundamentally a photospheric process.

The sequence lifetime of loops listed in Table \ref{tab:1} includes the time from the beginning of flux convergence, and until the end of divergence. In most cases, before the loop body appears, an intensity enhancement near its footpoints can be observed. Similarly, often when the loop body disappears, some intensity enhancement can still be seen near the loop footpoints. Also, in most cases, several individual loops appear and disappear, and loop decay and re-brightening can be observed, making the lifetime of the loop system look longer. When we track individual coronal loops in each sequence, the loop lifetime ranges from six minutes to 60 minutes, while as listed in Table \ref{tab:1}, the lifetime of principal loops ranges from 30 minutes to 120 minutes.

Nonetheless, the observations in general may have longer time scales during loop evolution than in the numerical simulations for the following reasons:
\begin{itemize}
    \item The numerical domain is relatively small, and the loops have shorter lengths.
    \item The convection zone does not cover deep layers, which limits the amplitudes and length and time-scales of any supergranular motions in the models.
    \item Due to resolution and time-scales from SDO/AIA and HMI, we may be biased towards long-lived loops and larger network (supergranular) patches. Note that we are also limited by the sensitivity of the HMI. 
\end{itemize}
As a result, the time-scales are much shorter in the simulations. But one would expect that longer loops attached to supergranular structures will have longer time-scales.

Despite this limitation, the Bifrost-enhanced network simulation allows interpretation of observations in which loop brightening is caused by photospheric magnetic flux convergence at loop footpoints. In the chosen synthetic loop, both footpoints experience episodes of convergence, but the flux convergence in patch (footpoint~1) consistently dominates and correlates with the loop luminosity. The horizontal flows maps at $\tau_{500}\!\approx\!1$ show enhanced flow convergence—and, to a lesser extent enhanced flow vorticity—preceding the increase of the volume-integrated emissivity \ion{Fe}{IX} in the loop field-line bundle. The cross-correlation between unsigned vertical flux and synthetic 171 \AA\ luminosity peaks when only pixels with fields above 200 G are included, consistent with the ``critical field" condition observed for AIA bright loops. 

We note that shear also tends to increase, particularly during the early stages of flux convergence. This suggests that convergence leads to enhanced shear among the magnetic flux elements, and that the increased shear may itself be indirectly responsible for the loop heating. In either case, flux convergence and shear appear to be closely linked processes, with convergence leading to shearing and thus being directly or indirectly responsible for the loop heating.

Previous studies have proposed potential mechanisms of coronal loop heating, including wave heating and nanoflares. These theories may work together with flux convergence and shear to drive loop heating. Flux convergence and shear may facilitate increased braiding and thus an increased rate of nanoflares. 
The convergence and the shearing component of the vertical Poynting flux strengthen at both footpoints. Co–temporal enhancements in $j^{2}/\rho$ and in the inverse characteristic length proxy $j/b$ mark enhanced current dissipation and topological reconfiguration (reconnection), with the volume-integrated emissivity peak lagging by about $150$~s. This temporal ordering-photospheric convergence leading to an increased shear–driven Poynting flux, followed by current–sheet formation, and subsequently enhanced coronal volume-integrated emissivity—supports a causal energy–transfer chain consistent with a convergence–driven increase in nanoflare–like reconnection events or an increase in chromospheric thermodynamics and spicule formation \citep{bose23}. The proposed idea is consistent with small-angle magnetic reconnection heating in closed inert loops \citep{parker83a,parker88}, for which some observational support was recently provided by \cite{moor24}. 
We note a caveat that the role of waves in the simulations may be underestimated due to insufficient spatial resolution \citep{asga12,vanb14}. In general, stronger magnetic fields associated with flux convergence are also expected to generate stronger waves. 

The increase in the heating near the converging-flux loop foot is also consistent with the enhanced Poynting flux compared to those in the neighboring quiet Sun regions having a weaker magnetic field \citep[e.g.,][]{chen14}. The upward energy input in loops is plausibly driven by the advection of the photospheric magnetic field, i.e., by the horizontal convective motions. In this scenario, enhanced Poynting flux at one loop end propagates to the other end, giving a short delay in heating at the other foot. The upward-directed Poynting flux would increase energy input, thus heating the coronal plasma and increasing the gas pressure due to the evaporation-driven upflows. While \cite{chen14} proposed the above heating mechanism primarily for an emerging flux loop system, it can also be valid after emergence has ceased, and as flux convergence continues to drive heating.

Note that a very accurate estimate of velocity and magnetic field is required to reliably estimate Poynting flux in observations \citep{wels15}. Due to the large noise level in the Bx and By components of the magnetic field measured by HMI \citep{sche12}, particularly for weak magnetic fields in quiet solar regions \citep{apar25}, a measurement of the Poynting flux is unreliable and has not been performed here. We rely on the simulation results for this aspect.

The asymmetry in footpoint energy injection seen in the simulation provides a natural explanation for the often one–sided heating signature inferred in several observed loops (see Table \ref{tab:1}). Still, the shear reaches roughly twice the amplitude and peaks closer in time to the \ion{Fe}{IX} maximum intensity at the dominant footpoint. Thus, heating in loops may happen due to an increased rate of nanoflares, and these nanoflares may concentrate on one side of the loop, near the converging flux loop-leg. This perhaps also indicates the heating in these loops to be asymmetric, due to propagating from one (converging) end to the other by thermal conduction. Most solar coronal loops are reported to display symmetric heating \citep[e.g.,][]{real14}.  However, the cadence of even 12 seconds of the AIA data cannot resolve the propagating heating along loops. A future investigation of loop heating using higher cadence data from Solar Orbiter/EUI \citep{mull20,roch20}, Hi-C \citep{koba14,rach19,tiw25}, or from other future higher spatial and temporal resolution instruments such as MUSE \citep{depo22,cheu22} and EUVST \citep{shim19}, see also the proposal of \cite{ster23}, may be more suitable to resolve the propagating brightening along the loop in asymmetric loops. Asymmetric loop heating has already been reported in \ion{Fe}{ix}/X emission, but for much shorter and smaller loops using high spatial and temporal resolution data from Hi-C 2.1 \citep{tiw19}, and Solar Orbiter/EUI \citep{tiw22}.

Figure \ref{fig:drawing} shows the idea for why nanoflare heating in a loop’s body increases as the magnetic field in either foot gets more tightly packed together. The cartoon is also a way to illustrate the closely similar conjecture that the rate of fine-scale braiding of the loop’s field lines by convective mixing in the feet increases as either foot becomes more compact. We note that the cartoon does not account for the possible role of shear and waves in heating, which may also be influenced by magnetic flux convergence.

The converging flux model of E. Priest for heating of coronal X-ray bright points [also known as coronal bright points, CBPs] suggests that magnetic flux convergence and cancellation could directly deposit the magnetic energy in the form of thermal energy in the corona \citep{prie94}. \cite{madj23} found observational evidence for this scenario and reported many coronal bright points forming either from magnetic flux emergence or by magnetic flux coalescence, by the chance encounter of close-by magnetic flux concentrations of opposite polarities. They further provided a range of the photospheric magnetic flux in bright and faint loops in CBPs, that comes out to be from 3 $\times$ 10$^{18}$ Mx to 1.8 $\times$ 10$^{20}$ Mx. The estimated magnetic flux in our 10 QR coronal loops during their peak luminosity time, ranges between $1\times10^{19}$ Mx and 7.2$\times 10^{19}$ Mx, consistent with the above numbers. 
It should be noted, however, that many coronal X-ray/EUV bright points are formed as a result of emerging flux \citep{madj19,tiw22}. In this work, we specifically avoided cases of loop brightening associated with emerging-flux regions in the photosphere.

Coronal loop heating associated with magnetic flux cancellation has been reported in several observational and theoretical studies \citep{tiw14,tiw21,chit17_sunrise,prie18,synt20,park20}. However, it has also been argued that the degree of convective freedom and the magnetic field strength at both loop footpoints primarily determine the amount of loop heating \citep{tiw17,tiw21}, implying that mixed-polarity magnetic fields are not necessarily required for coronal loop heating \citep{tiw21}. In any case, coronal loop heating associated with magnetic flux convergence has not previously been reported. In the coronal loops studied here, we do not find obvious mixed-polarity magnetic flux patches at the loop footpoints, suggesting that flux cancellation is likely not the dominant process at work. Nonetheless, future high-resolution magnetograms, such as those from DKIST, may reveal fine-scale mixed-polarity flux within regions that currently appear unipolar.


It is worth noting that magnetic fields on observable scales may not directly control coronal heating. Instead, they may primarily set the Alfvén speed and magnetic topology without determining the plasma heating rates \citep{stur99}. In particular, \cite{judg24} found no clear relationship between EUV coronal heating and chromospheric magnetic field quantities, such as the Poynting flux, and argued that “the corona is a self-regulating forced system.” In this view, heating depends primarily on the thermodynamic state of the corona rather than on boundary conditions. Nonetheless, our results suggest the opposite, thereby opening avenues for more detailed investigation. 

Note that there may be widespread weak and small-scale convergence occurring globally around the Sun, without making outstanding loop heating visible in \FeIX\ emissions. For loops exhibiting convergence but no noticeable heating in 171 \AA\ images, it is possible that heating occurred at lower temperatures, or that convergence is not the primary driver. Further detailed investigations are needed to explore these possibilities.

\section{Conclusions}\label{sec:conclusions}

Using SDO/AIA and SDO/HMI observations, along with a radiative Bifrost 3D MHD simulation, we have demonstrated that the convergence of unipolar magnetic flux patches — at (at least) one loop footpoint — directly correlates with coronal loop heating, evidenced by the loop's EUV emission. Furthermore, the divergence of magnetic flux at that footpoint tracks the dimming of the loop in coronal EUV emission. Although in the models the time-scales are shorter. 

We find that the correlation between AIA 171 Å luminosity and HMI B$_{los}$ strengthens significantly when the foot clump's area averaged magnetic field exceeds a critical threshold of approximately 200–500 G for different loops. While previous studies have reported that many of the hottest coronal loops in ARs are rooted in mixed-polarity flux at one or both footpoints \citep[][and references therein]{chit20,tiw21} — suggesting that magnetic reconnection in the lower solar atmosphere, often accompanied by magnetic flux cancellation, plays a key role in heating — our findings reveal a new aspect. Specifically, it has not been reported previously, for either active region or quiet region loops, that unipolar magnetic flux convergence at a loop footpoint alone can increase coronal loop heating. Future higher-sensitivity magnetograms may reveal mixed polarity in regions that appear unipolar in HMI data. For example, a comparison of HMI and ViSP/DKIST magnetograms in \cite{pane26_dkist} reveals both weaker and stronger fields in ViSP than in HMI, as well as additional minority-polarity flux. Similarly, \cite{nobr24} found that SST/CRISP resolved an emerging bipole much more clearly than SDO/HMI. If more advanced magnetographs reveal increasing amounts of fine-scale minority-polarity flux with increasing flux convergence, this could provide an alternative explanation for our results.

While magnetic flux convergence may not be the sole or dominant heating mechanism for all coronal loops in the solar atmosphere, our observational and modeling results suggest that it could make an important contribution to coronal heating and may help initiate the mechanisms responsible for it. Flux-convergence-driven heating may also operate in other solar environments, including active regions, network and plage areas, quiet-Sun regions, and coronal holes. However, more comprehensive and detailed investigations are needed to rigorously evaluate and confirm its broader role in solar coronal heating.

We propose that, as illustrated schematically in Figure \ref{fig:drawing} and supported by the simulated emissivity distributions (Figures \ref{fig:sim_example_loop} \& \ref{fig:sim_loop_3D_lines}), an enhanced occurrence rate of nanoflares within the loop volume, particularly concentrated near the converging footpoint, may play a central role in driving the heating of these structures. While Figure \ref{fig:drawing} is intended as a conceptual illustration, the simulations indicate a clear correlation between elevated values of current density squared (J²) and the ratio J/B with localized heating, as well as with regions of magnetic shear and flux convergence. This relationship represents the key physical connection underlying our interpretation. A caveat is that waves may also be affected by magnetic flux convergence. Since the simulations do not adequately capture the wave field, we cannot rule out their potential contribution to heating these loops.

The heating of loops rooted in the converging flux footpoint is also consistent with the enhanced Poynting flux expected in these regions, as verified in our simulation loop, compared to the surrounding quiet Sun areas with weaker magnetic fields.

Future work should extend this analysis to active-region loops and other solar environments, with the aim of systematically evaluating this mechanism and quantifying its contribution to solar coronal heating.
\\

We thank the referee for their positive remarks and constructive suggestions for improving the manuscript. S.K.T., S. B., N.K.P., J.M.S., and R.L.M gratefully acknowledge support from NASA HGI (80NSSC21K0520) and/or HSR (80NSSC23K0093) grants. S.K.T. also acknowledges support from NSF AAG award (no. 2307505), AGS/SHINE award (no. 2230633), NASA contract NNM07AA01C (Hinode), NASA HGI award (80NSSC24K0551), and ARC-CREST (NASA Cooperative Agreement 80NSSC23M0230).  
N.K.P.'s research was also supported by NASA contract NNG04EA00C (SDO/AIA), NASA's  HCSI (80NSSC25K7028) grant, and HSR grant (80NSSC24K0258). AIA and HMI are instruments onboard the Solar Dynamics Observatory, a mission for NASA's Living With a Star program. The AIA and HMI data are courtesy of NASA/SDO and the AIA and HMI science teams. J.M.S and S.B. acknowledge support by NASA contract NNG09FA40C (IRIS) and 80GSFC21C0011 (MUSE), NASA grant 80NSSC26K0018, NSF grants AGS2532363 and AGS2532187. S.B. also acknowledges support from the Research Council of Norway (project number: 360813) and through its Centres of Excellence scheme, project number 262622. B.D.P. acknowledges support by NASA contracts NNG09FA40C (IRIS) and 80GSFC21C0011 (MUSE).
The simulations were performed on resources provided by
Sigma2 - the National Infrastructure for High-Performance Computing and Data Storage in Norway, and on clusters, the Pleiades cluster, through the computing projects s1061, s2601 from the High-End Computing (HEC) division of NASA.
The authors acknowledge the use of the VAPOR visualization software (http://www.vapor.ucar.edu) to generate relevant graphics. This research has utilized NASA's Astrophysics Data System and the IDL SolarSoft and Python SunPy \citep[v7.0.4,][]{sunpy_package_cite} packages. CHIANTI is a collaborative project involving George Mason University, the University of Michigan (USA), the University of Cambridge (UK), and NASA Goddard Space Flight Center (USA).


\appendix

\section{Loops 4 and 7 from Table 1}

\begin{figure*}[h]
	\centering
	\includegraphics[trim=0cm 0cm 0cm 0cm,clip,width=\linewidth]{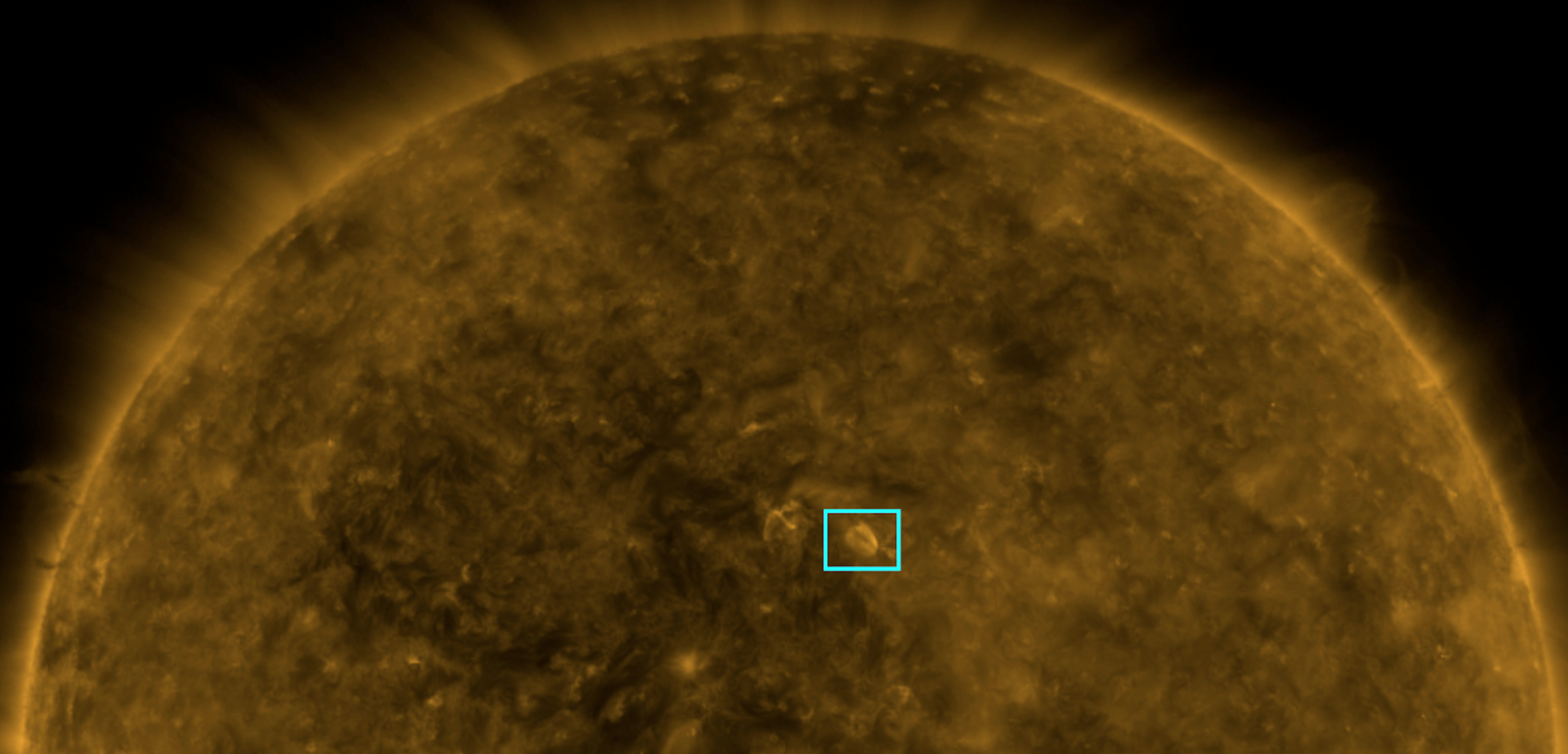}
	\includegraphics[trim=0cm 0cm 0cm 0cm,clip,width=\linewidth]{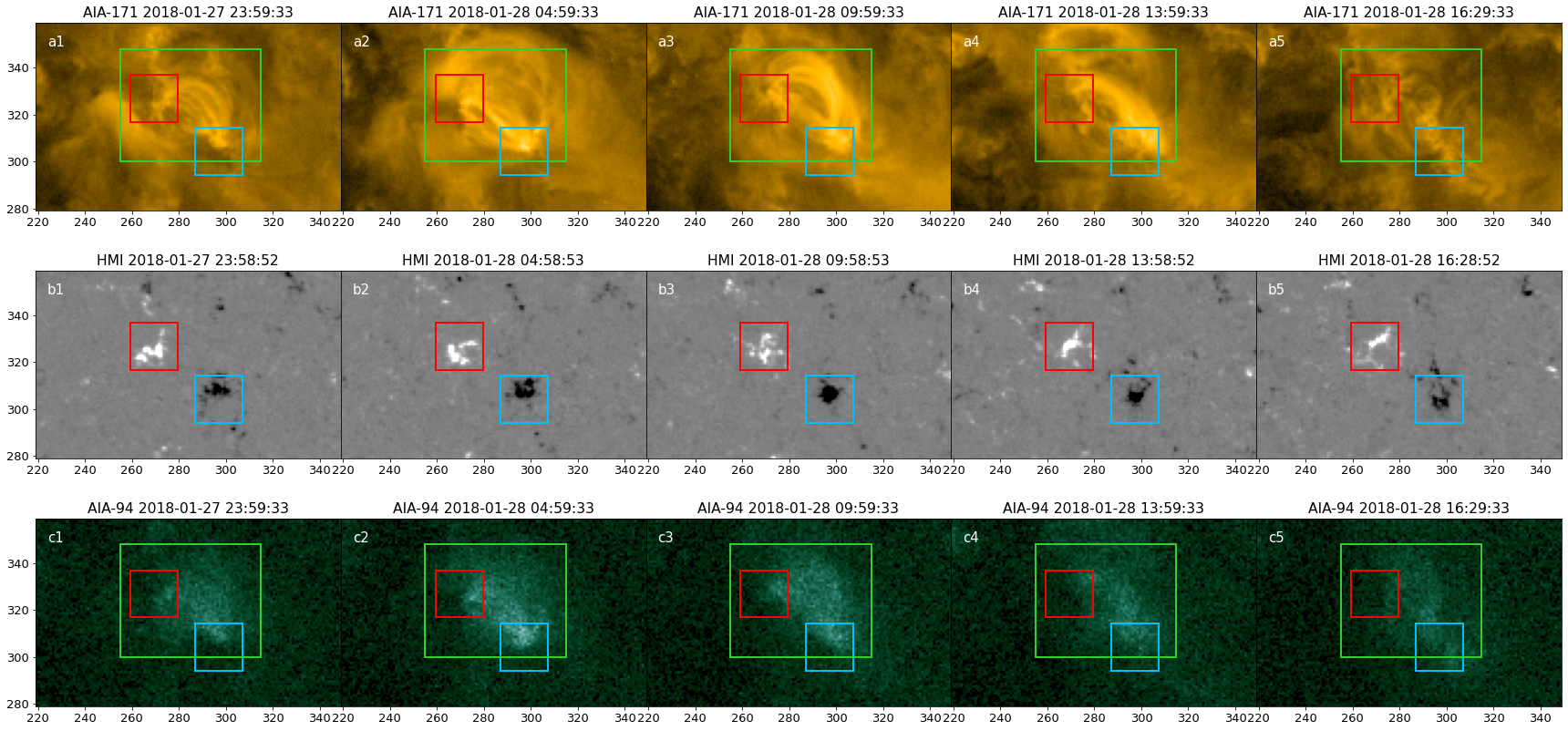}
	\caption{Same as Figure \ref{fig:1_context} but for loop 4 from Table \ref{tab:1}. In AIA 171 and 94 \AA\ images, a green large box outlines the area in which the loop intensities, plotted later in Figure \ref{fig:app1c_plots}, were calculated. An animation for $\sim$22 hours at a six-minute temporal cadence is available online (Movie3\_loop\_index4.mp4). The animation runs for 11 seconds.} 
	\label{fig:app1_context}
\end{figure*}

\begin{figure*}[h]
	\centering
    \includegraphics[trim=1cm 11.5cm 3cm 7cm,clip,width=\linewidth]{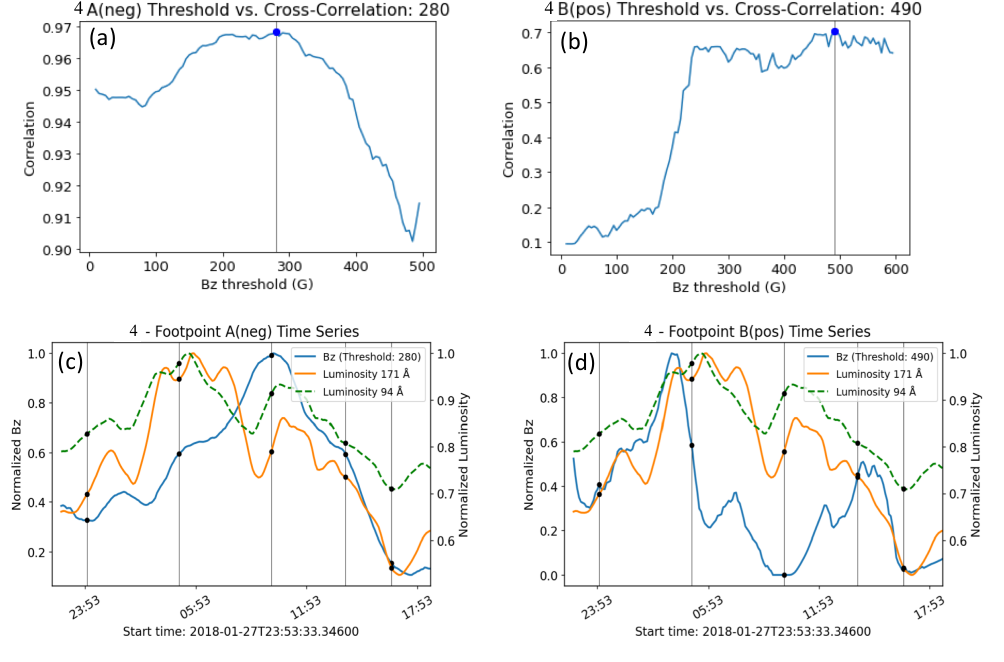}
	\caption{Same as Figure \ref{fig:2_plots} but for the example loop shown in Figure \ref{fig:app1_context}. The vertical five lines mark the timing 
    of five images shown in the last three rows of Figure \ref{fig:app1_context}. This is one of the clearest examples, as there is only a short delay between the onset of magnetic-flux convergence and the formation of the loop.} 
	\label{fig:app1c_plots}
\end{figure*}

We present here similar images and plots as Figures \ref{fig:1_context} and \ref{fig:2_plots}, but for two additional examples from Table \ref{tab:1}. Figure \ref{fig:app1_context} shows AIA images and magnetograms of loop 4. Figure \ref{fig:app1c_plots} presents corresponding plots.

\newpage

\begin{figure*}[h]
	\centering
	\includegraphics[trim=0cm 0cm 0cm 0cm,clip,width=0.91\linewidth]{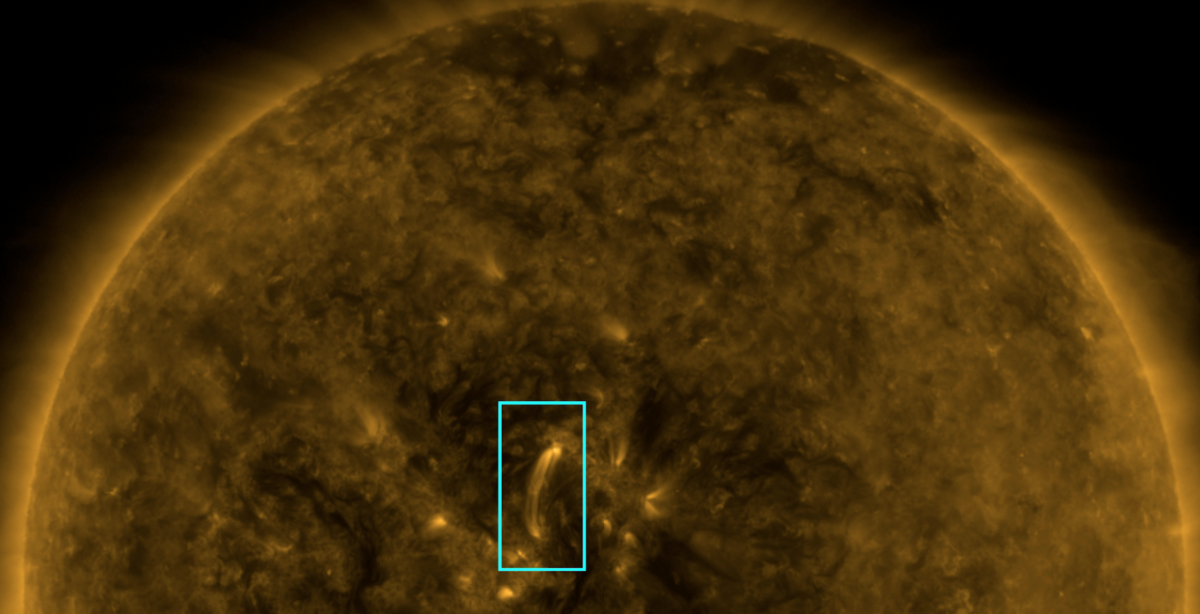}
	\includegraphics[trim=0cm 0cm 0cm 0cm,clip,width=0.91\linewidth]{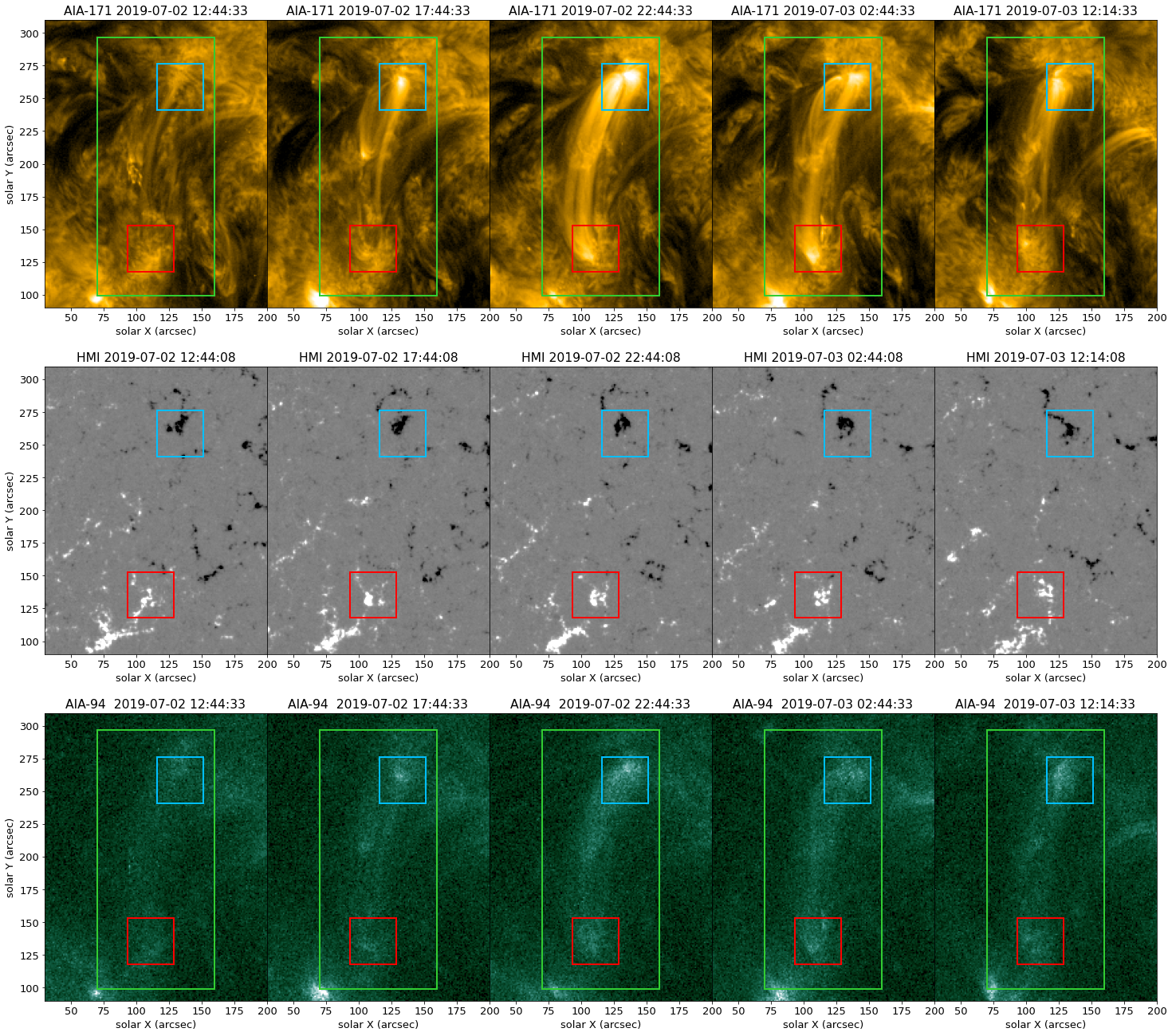}
	\caption{Same as Figure \ref{fig:1_context} but for loop 7 from Table \ref{tab:1}. In AIA 171 and 94 \AA\ images, a green large box outlines the area in which the loop intensities, plotted later in Figure \ref{fig:app2c_plots}, were calculated. An animation (Movie4\_loop\_index7.mp4) is available online. The animation contains AIA 171 and 94 Å images and HMI LOS magnetograms covering approximately 24 hours at a six-minute cadence. The animation runs for 12 seconds.} 
	\label{fig:app2_context}
\end{figure*}

\begin{figure*}[h]
	\centering
    \includegraphics[trim=1cm 12cm 3.5cm 7cm,clip,width=\linewidth]{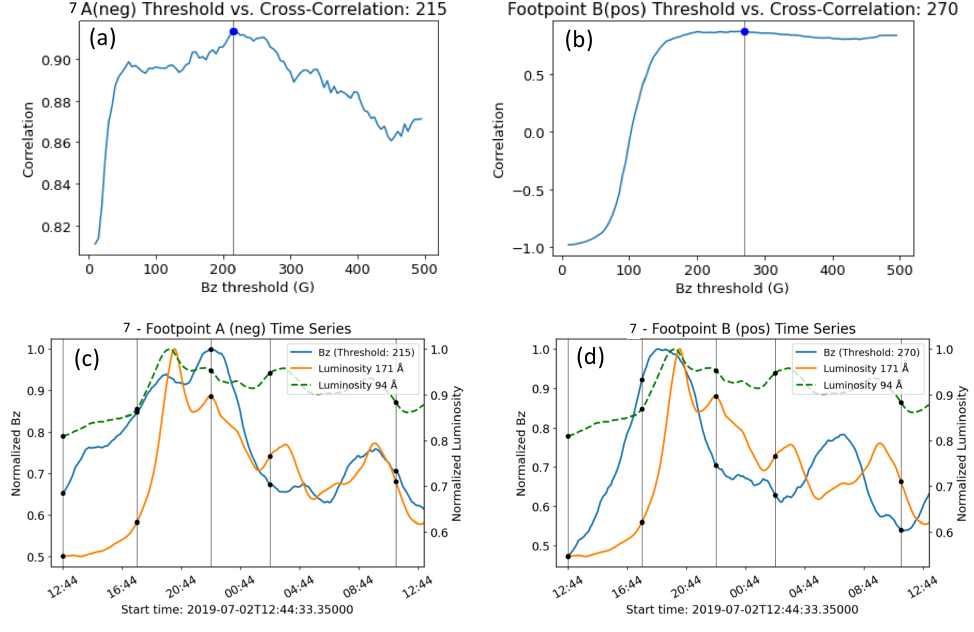}
	\caption{Same as Figure \ref{fig:2_plots} but for the example loop shown in Figure \ref{fig:app2_context}. The vertical lines mark the timing of five images shown in the last three rows of Figure \ref{fig:app2_context}. } 
	\label{fig:app2c_plots}
\end{figure*}

Figure \ref{fig:app2_context} shows AIA images and magnetograms of loop 7. Figure \ref{fig:app2c_plots} presents corresponding plots. 

\newpage
\section{Determining the critical magnetic field strength from the Bifrost simulation}

We compute a cross-correlation between the magnetic field threshold and the intensity at each footpoint within the selected box. The Pearson correlation for each magnetic field threshold is shown in the middle row. One can see that the optimum value is between 140-200~G. This also provides some confidence that using 200~G as the observation threshold is a good choice for finding correlations. Therefore, we have selected 200~G also for the simulations. 

\label{appendix_sim}

\begin{figure*}
\centering
    \includegraphics[height=18cm, width=12cm]{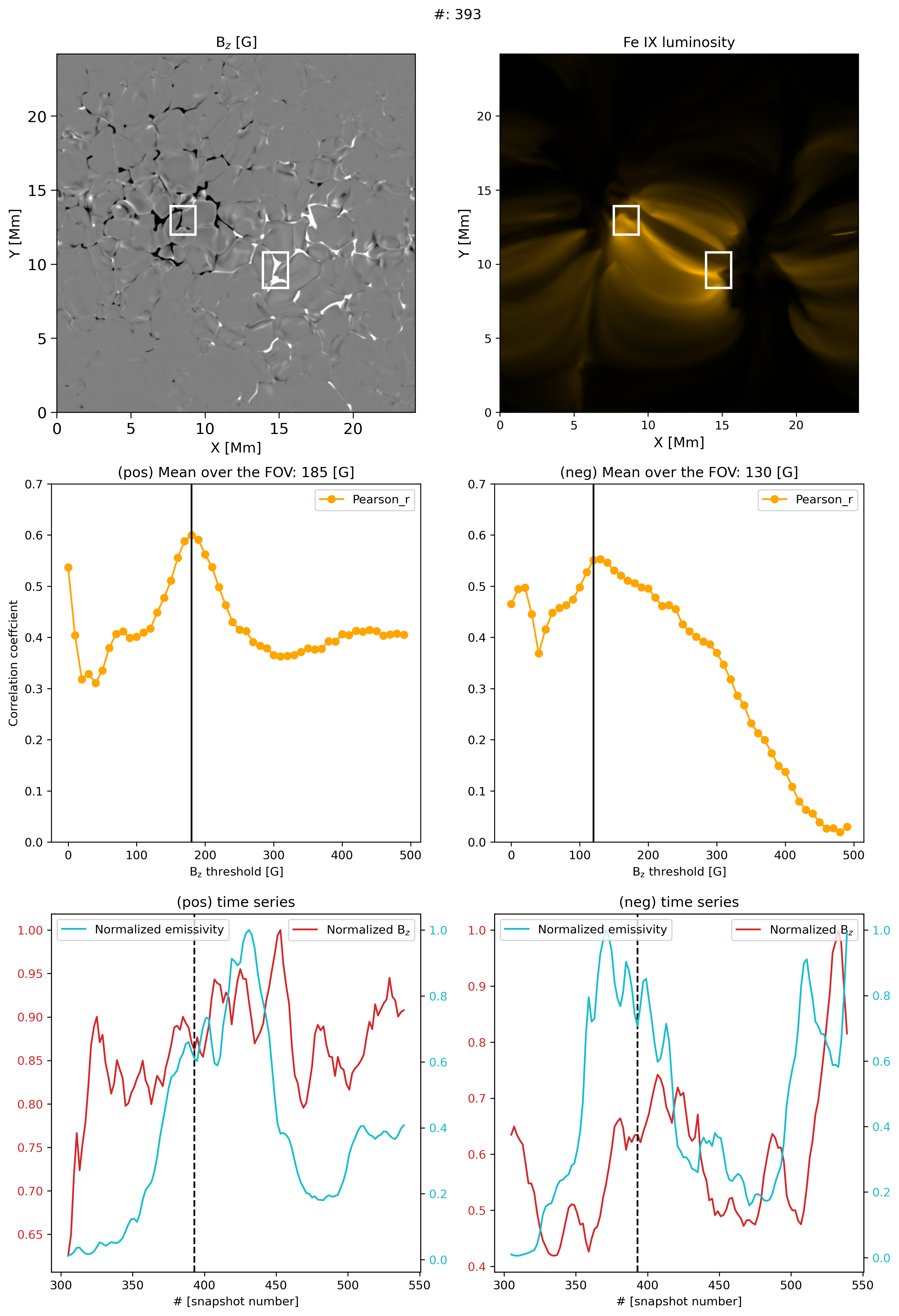} 
    \caption{Determination of the critical magnetic field strength from the enhanced network simulation. \textbf{Top row:} the left panel shows the $B_z$ map, and the right panel displays the synthetic \ion{Fe}{ix} emission map; two boxes outline the footpoints of the selected loop. \textbf{Middle row:} correlation between $B_z$ and luminosity at each loop footpoint. \textbf{Bottom row:} normalized emissivity and $B_z$ profiles at the location identified in the middle row (vertical lines), corresponding to the strongest correlation. An animation (Movie9.mp4) is available online. The animation displays every other simulation frame and runs for seven seconds.}
    \label{fig_appendix:critical_field_sim}
\end{figure*}

\end{document}